\documentclass[journal]{IEEEtran}

\usepackage{epsfig}
\usepackage{amssymb}
\usepackage{amsthm}
\usepackage{graphicx}
\usepackage{xcolor}
\usepackage{boldline,multirow}
\usepackage{hyperref}
\usepackage{amsmath}
\usepackage{amsthm}

\newtheorem{assumption}{Assumption}
\newtheorem{proposition}{Proposition}

\usepackage{pifont}

\begin{document}

\title{Tube-Based Robust Control Strategy for Vision-Guided Autonomous Vehicles}

\author{Der-Hau Lee
\thanks{The author was affiliated with the Department of Electrophysics,  National Yang Ming Chiao Tung University, Hsinchu 300, Taiwan. e-mail: derhaulee@gmail.com.}
% <-this % stops a space
}

% make the title area
\maketitle

\begin{abstract}
A robust control strategy for autonomous vehicles can improve system stability, enhance riding comfort, and prevent driving accidents. This paper presents a novel interpolation-tube-based constrained iterative linear quadratic regulator (itube-CILQR) algorithm for autonomous computer-vision-based vehicle lane-keeping. The goal of the algorithm is to enhance robustness during high-speed cornering on tight turns. Compared with standard tube-based approaches, the proposed itube-CILQR algorithm reduces system conservatism and exhibits higher computational speed. Numerical simulations and vision-based experiments were conducted to examine the feasibility of using the proposed algorithm for controlling autonomous vehicles. The results indicated that the proposed algorithm achieved superior vehicle lane-keeping performance to variational CILQR-based methods and model predictive control (MPC) approaches involving the use of a classical interior-point optimizer. Specifically, itube-CILQR required an average runtime of 3.45 ms to generate a control signal for guiding a self-driving vehicle. By comparison, itube-MPC typically required a 4.32 times longer computation time to complete the same task. Moreover, the influence of conservatism on system behavior was investigated by exploring the variations in the interpolation variables derived using the proposed itube-CILQR algorithm during lane-keeping maneuvers.

\end{abstract}


\section{Introduction}
\IEEEPARstart{I} {}n various scientific, engineering, and real-world tasks, designing an appropriate control law for a given dynamical system is generally necessary. Advances in computing have led to the increasing use of optimization-based model predictive control (MPC) techniques to guide working systems. MPC techniques can optimize forecasts of a constrained system's behavior over a finite time horizon window on the basis of a state-space model with an associated performance-based cost function. An MPC controller generates a control law by solving the associated regulation problem, which is formulated as an optimal control problem. MPC controllers are more flexible than classic linear quadratic regulator (LQR) controllers for solving complex control problems with various constraints. However, for these problems, LQR controllers often cannot provide an analytic solution. MPC algorithms can also be used in feedback control by repeatedly solving the open-loop optimal control problem. The resulting state feedback law can then be used to generate adjustable inputs for regulating a system \cite{Raw00,Svr23}. Nevertheless, uncertainty, oversimplified modeling conditions, incorrect model assumptions, and external disturbances to the state or control variables can lead to a decrease in the forecasting accuracy of a classical MPC algorithm. To enhance the robustness of classical MPC, researchers have proposed tube-based MPC (tube-MPC) approaches \cite{Raw17}. Tube-MPC requires all trajectories of the system to lie in a bounded neighborhood (a tube) that satisfies all system constraints. The cross section of the tube is determined by introducing uncertainty to a suitably constrained nominal (noise-free) system. To avoid the generation of an excessively large tube, a feedback control policy $u$ is applied to the system, where $u = \bar u + K\left( {x - \bar x} \right)$ and $\bar x$ and $\bar u$ are the nominal state and control variables, respectively. The LQR feedback gain $K$ is selected to regulate the deviation between the actual and nominal states such that this deviation converges to zero. Tube-MPC approaches ensure robustness, stability, and constraint fulfillment; therefore, they have been widely used in recent automotive \cite{Gao14,Mat19,Kha21,Wan22,Lu25} and robotics \cite{Gon12,Kog20,Dav11,Wil18,And25} applications.

The authors of \cite{Gao14} developed a nonlinear bicycle model with force inputs and applied it in a tube-MPC controller to maneuver a vehicle in a static-obstacle-avoidance task. In several trials of the controlled system under random external noise, the vehicle trajectories produced by the tube-MPC controller were considerably tighter (i.e., less scattered) than those produced by a classical MPC controller. Furthermore, the authors of \cite{Mat19} employed a linear single-track model in a vehicle lane-keeping application. Tube-MPC was used to ensure ride comfort and driving stability if model deviations occurred. In \cite{Kha21}, a unified framework was proposed for a self-driving vehicle, which was navigated by a tube-MPC controller to robustly and safely follow a reference trajectory generated by a Frenet frame planner. This framework allowed the vehicle to avoid dynamic obstacles. The control component of the framework accounted for not only external state and control disturbances but also internal model uncertainty. Online estimation of the size of state disturbance was achieved through Gaussian process regression. In \cite{Wan22}, a tube-MPC strategy was proposed and applied to achieve path-following for four-wheel, independent-drive automated vehicles. This model exhibited time-varying uncertainty attributable to tire cornering stiffness. However, the control performance of conservative (restrictive) tube-MPC was inferior to that of classical MPC when white noise was added to the state variables. The authors of \cite{Lu25} proposed a stochastic tube-MPC method subject to constraints with  probability distributions. A dynamic feedback gain mechanism was designed on the basis of covariance steering theory, and tightened constraints were determined using the concept of probabilistic reachable sets. The nominal constraint region was larger for stochastic tube-MPC than for tube-MPC, indicating that the stochastic approach was an improvement on the excessively conservative tube-MPC method.

In \cite{Gon12}, the tube-MPC technique was used to conduct trajectory tracking for a mobile robot with longitudinal slip. The investigated system was decomposed into time-invariant and time-varying components, and a specific set of control gains was selected to compensate for the corresponding system dynamics such that the mismatch between the actual and nominal systems approached zero. Instead of using fixed feedback gains in tube-MPC,  the authors of \cite{Kog20} employed an interpolation method to conduct online fusion of multiple tubes generated using different control gains. This approach reduced mismatches between predictions and the ground truth. The approach guarantees robust stability, feasible trajectories, and constraint satisfaction at a marginal increase in computational cost. Moreover, it provides a wider range of feasible region variables compared with those obtained in an offline approach with fixed constraint bounds. In \cite{Dav11} and \cite{Wil18}, feedback from a real-world system was considered to improve tube-MPC. In tube-MPC, an idealized nominal state is simulated without considering any environmental response. Thus, the performance of the algorithm is completely reliant on the tracking ability of the feedback law for system control. The authors of \cite{Dav11} used both the actual and nominal states as inputs to the nominal controller and compared the resulting solutions. The superior solution was then selected as the control law for guiding the system. The authors of \cite{Wil18} proposed an algorithmic framework for model predictive path integral control. This framework uses a tube-MPC technique that is a relaxed version of the mechanism employed in \cite{Dav11}. The algorithm developed in \cite{Wil18} overcomes the inherent problem of low robustness in sampling-based MPC. In \cite{And25}, a rapid scaled-symmetric optimization algorithm based on the alternating direction method of multipliers algorithm was developed to enhance the computational efficiency of embedded tube-MPC for a high-order unmanned aerial vehicle (UAV) system. The proposed algorithm exhibited strong performance in solving the trajectory tracking problem for a tiltrotor UAV carrying a suspended load.

Motivated by the aforementioned studies, the present study developed an interpolation-tube-based constrained iterative linear quadratic regulator (itube-CILQR) algorithm for vehicle control \cite{Che17, Che19,Liu24}. The CILQR algorithm is a version of the MPC algorithm for solving optimal control problems. The CILQR algorithm uses barrier functions to handle system constraints in an iterative linear quadratic regulator framework \cite{Tas14}. These constraints are included in the control objective to convert the original control problem into an effectively unconstrained problem.  This barrier-function-based strategy facilitates rapid optimization and enables the generation of smooth solution trajectories. However, robustness is not guaranteed when the classical CILQR algorithm is applied to systems with disturbances. Therefore, we adapted several tube-MPC techniques presented in \cite{Raw17,Kog20,Dav11} to enhance the robustness of the CILQR algorithm under disturbances. The features of the proposed itube-CILQR algorithm are as follows. First, it computes tightened constraints by using an outer-bounding tube with varying disturbance bounds \cite{Raw17}. Second, it generates a control law that can incorporate environmental feedback to maneuver the system \cite{Dav11}. Third, it employs an interpolation approach that incorporates adjustable variables to fuse multiple tubes and produce less conservative tightened constraints \cite{Kog20}. Reducing conservatism is crucial because conservatism often degrades controller performance in standard tube-based approaches \cite{Wan22, Lu25}. To demonstrate its feasibility, the proposed itube-CILQR algorithm was used to solve the optimal control problem for a vehicle lane-keeping task.  

Handling road curvature is key challenge in vehicle lane-keeping control. Technically, the standard single-track dynamics model becomes nonlinear when a curvature term is included \cite{Raj11, Xu20}. Hence, solving this problem requires computationally expensive nonlinear optimization, which is unsuitable for the real-time demands of many applications. Accordingly, we did not solve the nonlinear lane-keeping control problem directly, as was the case in previous studies \cite{Xu20, Get24}. Instead, we treated the curvature term as an external disturbance source and then applied the itube-CILQR algorithm to robustly solve the corresponding control problem. In our previous work \cite{Lee24}, we developed an efficient vision-based system that uses individual front-facing camera images to perceive a vehicle's surrounding environment. This system can predict the ego vehicle's heading angle and detect the ego-lane boundary. Validation in The Open Racing Car Simulator (TORCS) \cite{Wym13} demonstrated the aforementioned system to exhibit promising performance. This platform, including the visual perception system and TORCS environment, enables real-time simulated control of vehicles by vision-guided lane-keeping algorithms. Therefore, the platform was used in this study to evaluate the performance of the proposed algorithm. The main contributions of this study are as follows:
\begin{enumerate}
  \item The study developed the itube-CILQR control algorithm, which is based on the CILQR algorithm and tube-based techniques, to ensure the robustness and stability of real-time lane-keeping control for autonomous vehicles on roads with large curvatures.
  \item The study validated the performance of the proposed itube-CILQR algorithm in numerical simulations and vision-based autonomous driving experiments within the realistic TORCS simulator.
\end{enumerate}

The remainder of this paper is organized as follows. The adopted lane-keeping dynamics model is introduced in Section II. The MPC schemes for the lane-keeping problem are detailed in Section III. The proposed itube-CILQR algorithm is described in Section IV. The experimental results are discussed in Section V. Finally, the study conclusion is provided in Section VI.

\section{Lane-Keeping Dynamics Model}
This study used a bicycle model to develop a representation of vehicle lane-keeping dynamics \cite{Raj11}. The corresponding lateral control model is defined in terms of the distance of the ego vehicle from the lane centerline ($\Delta$), the orientation error of the ego vehicle with respect to the road direction ($\theta$), and the derivatives of these factors (i.e., ${\dot \Delta }$ and ${\dot \theta }$, respectively). For a controller designed to ensure automatic lane-keeping by a vehicle, the vehicle speed in the longitudinal direction is denoted as $v_x$, the front-wheel steering angle is denoted as $\delta$, and the road curvature is denoted as $\kappa$. Assuming that the lateral force of a tire is proportional to its slip angle, the  discrete-time dynamics model for vehicle lane-keeping control can be expressed as follows \cite{Xu20, Lee19}:
\begin{equation}
{\bf x}_{i + 1}  \equiv {\bf f}\left( {{\bf x}_i ,{\bf u}_i } \right)  = {\bf Ax}_i  + {\bf Bu}_i+ {\bf w},
\end{equation}
where the state vector and control input are defined as ${\bf x}  = \left[ {\begin{array}{*{20}c}
   {x_0 } & {x_1 } & {x_2 } & {x_3 }  \\
\end{array}} \right]^T  \equiv  \left[ {\begin{array}{*{20}c}
   \Delta  & {\dot \Delta } & \theta  & {\dot \theta }  \\
\end{array}} \right]^T$ and ${\bf u} \equiv  \left[  \delta  \right]$, respectively. The  term ${\bf w}$, which includes $\kappa$, is considered to represent an external additive disturbance. The model matrices $\bf A$, $\bf B$, and $\bf w$ have the following form: 
\[
\begin{array}{l}
 {\bf A} = \left[ {\begin{array}{*{20}c}
   {a _{11} } & {a _{12} } & 0 & 0  \\
   0 & {a _{22} } & {a _{23} } & {a _{24} }  \\
   0 & 0 & {a _{33} } & {a _{34} }  \\
   0 & {a _{42} } & {a _{43} } & {a _{44} }  \\
\end{array}} \right], \\ 
 {\bf B} = \left[ {\begin{array}{*{20}c}
   0  \\
   {b _1 }  \\
   0  \\
   {b _2 }  \\
\end{array}} \right],\quad {\bf w} = \kappa\left[ {\begin{array}{*{20}c}
   0  \\
   {c _1 }  \\
   0  \\
   {c _2 }  \\
\end{array}} \right] . \\ 
 \end{array}
\]
The corresponding matrix coefficients are expressed as follows:
\[
\begin{array}{l}
a _{11} = a _{33} = 1, \quad a _{12} = a _{34} = dt, \\
a _{22}  = 1 - \frac{{\left( {2C_{\alpha f}  + 2C_{\alpha r} } \right)dt}}{{mv_x }},\quad a _{23}  = \frac{{\left( {2C_{\alpha f}  + 2C_{\alpha r} } \right)dt}}{m}, \\
a _{24}  =  - \frac{{\left( {2l_f C_{\alpha f}  - 2l_r C_{\alpha r} } \right)dt}}{{mv_x }},\quad a _{42}  =  - \frac{{\left( {2l_f C_{\alpha f}  - 2l_r C_{\alpha r} } \right)dt}}{{I_z v_x }} , \\
a _{43}  = \frac{{\left( {2l_f C_{\alpha f}  - 2l_r C_{\alpha r} } \right)dt}}{{I_z }},\quad a _{44}  = 1 - \frac{{\left( {2l_f^2 C_{\alpha f}  + 2l_r^2 C_{\alpha r} } \right)dt}}{{I_z v_x }}, \\
b _1 = {\frac{{2C_{\alpha f} dt}}{m}} ,\quad b _2  = \frac{{2l_f C_{\alpha f} dt}}{{I_z }}, \\
c _1  =  - \frac{{\left( {2l_f C_{\alpha f}  - 2l_r C_{\alpha r} } \right)dt}}{m} - v_x^2 dt, \\
c _2  =  - \frac{{\left( {2l_f^2 C_{\alpha f}  + 2l_r^2 C_{\alpha r} } \right)dt}}{{I_z }}.\\
\end{array}
\]
Here, $dt$ is the sampling time; $m$ is the vehicle mass; $I_z$ is the moment of inertia along the vertical direction; $C_{\alpha f}$ and $C_{\alpha r}$ are the cornering stiffness values of the front and rear tires, respectively; and $l_f$ and $l_r$ are the distances from the center of gravity of the vehicle to the front and rear tires, respectively. The model and vehicle parameters selected in this study are listed in Table I.

As described in Sec. I, nearly optimal control inputs can be straightforwardly obtained by using the standard MPC scheme (Sec. V-C in \cite{Xu20}) to solve the constrained lane-keeping dynamics model (1). However, this strategy may have an excessive computational demand in practice because of the nonlinearity of the lane curvature $\kappa$, compromising driving safety under extreme driving conditions, such as high-speed cornering.

The objective of this study was to develop a rapid and robust tube-based controller for real-time automotive lane-keeping. To achieve this goal, (1) is considered to represent the actual system, which is subject to the stage constraints ${\bf x}_{i \ne N} \in \mathbb{X}$ and ${\bf u}_{i \ne N} \in \mathbb{U}$, the terminal constraint ${\bf x}_{N} \in \mathbb{T}$, and the disturbance ${\bf w} \in \mathbb{W}$. Therefore, the disturbance-free nominal system is defined as follows:
\begin{equation}
{\bf \bar x}_{i + 1}  \equiv {\bf \bar f}\left( {{\bf \bar x}_i ,{\bf \bar u}_i } \right) = {\bf A \bar x}_i  + {\bf B \bar u}_i ,
\end{equation}
where $\bf \bar x$ and $\bf \bar u$  are the associated nominal state and control vectors, respectively. Instead of solving the constrained optimization problem by employing the nonlinear model presented in (1), one can solve this problem by using the linear model expressed in (2) and process the ${\bf w}$ term through a tube-based approach, thus achieving a considerably lower computational cost with an acceptable reduction in performance. MPC schemes related to the considered lane-keeping problem are discussed in the following sections.

\begin{table}[!t]
\caption{Model and vehicle parameters used in this study}
\begin{center}
\begin{tabular}{llll}
\hline
$dt$ &  0.01 s & ${C_{\alpha f} }$  & 80000 N/rad  \\
$v_x$ & \{20.0, 22.2\} m/s  &  ${C_{\alpha r} }$ &  80000 N/rad  \\
$I_z$ &   2000 kgm$^{2}$ &  $l_f$  & 1.27 m \\
$m$ &  1150 kg & $l_r$ & 1.37 m  \\\hline
\end{tabular}
\end{center}
\end{table}

\section{MPC Schemes for the Lane-Keeping Problem}
This section introduces MPC schemes used to solve the lane-keeping problem. Details regarding the computation of tightened constraints for tube-based approaches are provided in the last part of this section. The operators $\oplus$ and $\ominus$ denote the Minkowski sum and the Pontryagin difference between sets, respectively.

\subsection{Nominal MPC scheme}
This section describes the nominal MPC algorithm. Here, the term \textquotedblleft nominal MPC" refers to MPC computation conducted using the nominal system model presented in (2), with no consideration of robustness. The optimization variables in this control problem are a sequence of states ${\bf X} \equiv \left\{ {{\bf x}_0 ,{\bf x}_1 ,...,{\bf x}_N } \right\}$ and  a control sequence ${\bf U} \equiv \left\{ {{\bf u}_0 ,{\bf u}_1 ,...,{\bf u}_{N - 1} } \right\}$. The associated optimization problem subject to constraints $\mathbb{X}$, $\mathbb{T}$, and $\mathbb{U}$ is formulated as follows (Problem 1):

\textit{Problem 1}:
\begin{subequations}
\begin{equation}
\mathop {\min }\limits_{{\bf X},{\bf U}} J = J_s  + J_t ,
\end{equation}
\begin{equation}
J_s  = \sum\limits_{i = 0}^{N - 1} {{\bf x}_i^T {\bf Qx}_i  + {\bf u}_i^T {\bf Ru}_i } ,
\end{equation}
\begin{equation}
J_t  = {\bf x}_N^T {\bf Px}_N 
\end{equation}
\end{subequations}
subject to
\begin{subequations}
\begin{equation}
{\bf x}_{i + 1}  = {\bf \bar f}\left( {{\bf  x}_i ,{\bf  u}_i } \right) ,\quad 0 \le i < N,
\end{equation}
\begin{equation}
{\bf x}_{i\ne N} \in \mathbb{X},
\end{equation}
\begin{equation}
{\bf x}_{N} \in \mathbb{T}, 
\end{equation}
\begin{equation}
{\bf u}_{i\ne N} \in \mathbb{U}, 
\end{equation}
\end{subequations}
where the total cost $J$ comprises the stage and terminal cost functions $J_s$ and $J_t$, respectively, and $N$ denotes the prediction horizon. The closed-loop asymptotic (exponential) stability of the aforementioned model predictive controller is ensured if the following assumptions are satisfied \cite{May00}:
\begin{assumption}
The convex sets $\mathbb{X}$ and $\mathbb{T}$ are closed,  $\mathbb{U}$ is compact, and $\mathbb{T}$ is contained in $\mathbb{X}$. In geometric terms, $\mathbb{X}$, $\mathbb{T}$, and $\mathbb{U}$ are polytopes that contain the origin in their interior.
\end{assumption}
\begin{assumption}
For a terminal control gain ${\bf K}_f$ and any ${\bf x} \in \mathbb{T}$, the control constraint is satisfied by the resulting control signal ${\bf u}_{f}$. 
\end{assumption}
\begin{assumption}
The set $\mathbb{T}$ is positively invariant under the control signal ${\bf u}_{f}$, and the maximal positively invariant set \cite{Kou16} is selected as constraint $\mathbb{T}$ in this study. 
\end{assumption}
\begin{assumption}
The costs $J_s$ and $J_t$ are quadratic [(3b) and (3c), respectively], and the weighting matrices ${\bf Q}$ and ${\bf R}$ are symmetric positive-definite. Moreover, $J_t$ is a local Lyapunov function, and the corresponding matrix ${\bf P}$ is the solution of the discrete-time algebraic Riccati equation, which is expressed as follows:
\begin{equation}
{\bf P} = {\bf A}^T {\bf PA} + {\bf Q} - {\bf A}^T {\bf PB}({\bf B}^T {\bf PB} + {\bf R})^{ - 1} {\bf B}^T {\bf PA},
\end{equation}
The terminal control gain ${\bf K}_f$ is defined as follows: 
\begin{equation}
{\bf K}_f =  - \left( {{\bf B}^T {\bf PB} + {\bf R}} \right)^{ - 1} {\bf B}^T {\bf PA}.
\end{equation}
\end{assumption}
Appropriate selection of $\mathbb{T}$, ${\bf P}$, and ${\bf K}_f$ can ensure that the following elementary closed-loop properties of Problem 1 are satisfied \cite{May00}:
\begin{proposition}
If Assumptions 1\textendash4 hold, the solution of Problem 1 yields the optimal control sequence and the associated optimal state sequence. The  model predictive control law ensures that the optimization problem remains feasible at all times and that the system is exponentially stable for all initial states in $\mathbb{X}$.
\end{proposition}

The control parameters  and constraints used in (3) and (4) are listed in Table II. The first component of the optimal control sequence of the MPC problem is selected to guide the ego vehicle. The relevant control law is expressed as follows:
\begin{equation}
u = u_d \equiv u^{*}_0 \left( {\bf x}_0 \right),
\end{equation}
where ${\bf x}_0$ represents the current actual states. Clearly, the solution obtained for Problem 1 by using the nominal model (2) is a rudimentary approximation of the true solution of the MPC problem for the actual system model (1). This nominal MPC scheme is valid at low speed or for small-curvature roads. In our previous study \cite{Lee24}, vehicles guided by this nominal MPC algorithm occasionally left the lane unintentionally when entering tight turns at high speed.

\begin{table}[!t]
\caption{Control parameter values and constraints}
\begin{center}
\begin{tabular}{llll}
\hline
$N$ & 30 & ${\bf Q}$ & diag[20, 1, 20, 1]   \\
$\Delta$  & [-2.0, 2.0] m & ${\bf R}$ & 60$\bf I$  \\
 $\dot \Delta $ &  [-9.0, 9.0] m/s & ${\bf S}$ & 50$\bf I$ \\
$\theta  $  & [-$\pi  /2$, $\pi  /2$]  rad &  $q_s$ & 100\\
$\dot \theta  $ & [-4.0, 4.0] rad/s & $q_u$ & 10 \\
$\delta$ &  [-$\pi  /6$, $\pi  /6$]  rad& $q_{l1}$ & 80 \\
$\kappa$ &  [-0.1, 0.1] 1/m& $q_{l2}$ & 20 \\
\hline
\end{tabular}
\end{center}
\end{table}

\subsection{Tube-MPC scheme}
The key concept of tube-MPC is that if the nominal state ${\bf \bar X} \equiv \left\{ {{\bf \bar x}_0 ,{\bf \bar x}_1 ,...,{\bf \bar x}_N } \right\}$ and the control sequence ${\bf \bar U} \equiv \left\{ {{\bf \bar u}_0 ,{\bf \bar u}_1 ,...,{\bf \bar u}_{N-1} } \right\}$ satisfy the tightened constraints $\mathbb{\bar X}$, $\mathbb{\bar T}$ (in $\mathbb{\bar X}$), and $\mathbb{\bar U}$, the state and control trajectories ($\bf X$ and $\bf U$, respectively) of the actual system also satisfy the original constraints $\mathbb{ X}$, $\mathbb{ T}$, and $\mathbb{ U}$ under the disturbance $\bf {w} \in \mathbb{W}$, with  the uncertainty set $\mathbb{S} = \mathbb{S}\left( \mathbb{ W} \right)$. Thus, all trajectories of the actual system lie in the tube of the tube-MPC framework, which guarantees system stability and  enhances system robustness \cite{Raw17}. Constraint tightening is achieved by using $\mathbb{\bar X} \triangleq \mathbb{X} \ominus \mathbb{S}$ and $\mathbb{\bar U}\triangleq \mathbb{U}  \ominus \bf K \mathbb{S}$ to bound the deviation of the actual state from the nominal state. The relevant computations are presented in Sec. III-D.

The corresponding tube-MPC problem subject to tightened constraints is formulated as follows (Problem 2):

\textit{Problem 2}:
\begin{subequations}
\begin{equation}
\mathop {\min }\limits_{{\bf \bar X},{\bf \bar U}} \bar J = \bar{J}_s  + \bar{J}_t ,
\end{equation}
\begin{equation}
\bar{J}_s  = \sum\limits_{i = 0}^{N - 1} {{\bf \bar x}_i^T {\bf Q \bar x}_i  + {\bf \bar u}_i^T {\bf R \bar u}_i } ,
\end{equation}
\begin{equation}
\bar{J}_t  = {\bf \bar x}_N^T {\bf P \bar x}_N 
\end{equation}
\end{subequations}
subject to
\begin{subequations}
\begin{equation}
{\bf\bar x}_{i + 1}  = {\bf \bar f}\left( {{\bf \bar x}_i ,{\bf \bar u}_i } \right) ,\quad 0 \le i < N,
\end{equation}
\begin{equation}
{\bf \bar x}_{i\ne N} \in \mathbb{\bar X},  
\end{equation}
\begin{equation}
{\bf \bar x}_{N} \in \mathbb{\bar T}, 
\end{equation}
\begin{equation}
{\bf \bar u}_{i\ne N} \in \mathbb{\bar U}.  
\end{equation}
\end{subequations}
The control law $u$ for the actual system contains two terms: i) the solution of the  MPC problem ($\bar u_{MPC}$) and ii) a tracking control term ($\bar u_{LQR}  = {\bf K}\left( {{\bf x} - {\bf \bar x}} \right)$) that compensates for the discrepancies between the nominal and actual states. The stabilizing LQR gain (6) is selected as the control gain ${\bf K}$ to attenuate disturbances.

The closed-loop properties of Problem 2 are ensured if the following assumptions
are satisfied \cite{May05}:
\begin{assumption}
The admissible disturbance $\bf {w} \in \mathbb{W}$; that is, the bounds of $\bf {w}$ are given. Here, $\bf {w}$ is compact and contains the origin. Moreover, the set $\mathbb{W}$ is sufficiently small for ensuring that $\mathbb{\bar X}$, $\mathbb{\bar T}$, and $\mathbb{\bar U}$ have an interior.
\end{assumption}
\begin{assumption}
The state transition matrix ${\bf A}_{K}  \equiv{\bf A}+{\bf B}{\bf K}$ is stable, and the terminal set $\mathbb{\bar T}$ is positively invariant under the control law $u$. 
\end{assumption}
\begin{assumption}
The costs $\bar{J}_s$ and $\bar{J}_t$ satisfy the usual axioms (e.g., Assumption 4).
\end{assumption}
Under the aforementioned assumptions, the following fundamental properties of the standard tube-MPC scheme (Problem 2) are guaranteed:
\begin{proposition}
If Assumptions 5\textendash7 hold, the solution of Problem 2 yields the optimal control sequence and the associated optimal state sequence for the nominal system. The control policy ensures that the states of the actual system are close to those of the nominal system. The actual system achieves constraint satisfaction, robust recursive feasibility, and robust stability if the initial nominal state is feasible.
\end{proposition}

To enable the nominal trajectory to respond to environmental changes, the authors of \cite{Dav11} compared two copies of a controller against each other: one was obtained from the nominal state ($u_n$), and the other was obtained from the actual  state ($u_a$). These controllers are expressed as follows:
\begin{equation}
\begin{split}
u_n \equiv u\left( {{\bf \bar x}_0} \right) &= \bar u_{MPC} \left( {{\bf \bar x}_0} \right) + \bar u_{LQR} \left( {{\bf \bar x}_0} \right) \\
&= \bar u_{MPC} \left( {{\bf \bar x}_0} \right) + {\bf K}\left( {{\bf x}_0 - {\bf \bar x}_0} \right)
\end{split}
\end{equation}
and
\begin{equation}
\begin{split}
u_a \equiv u\left( {\bf x}_0 \right) &= \bar u_{MPC} \left( {\bf x}_0 \right) + \bar u_{LQR} \left( {\bf x}_0 \right) \\
&= \bar u_{MPC} \left( {\bf x}_0 \right),
\end{split}
\end{equation}
where ${\bf \bar x}_0$ is the current nominal state and $\bar u_{LQR} \left( {\bf x}_0 \right)$ = 0. The solution with lower cost is selected as the superior and applied to control the actual system. In this study, we synthesize (10) and (11) to control the actual system. The proposed control policy ($u_p$) is expressed as follows:
\begin{equation}
\begin{split}
u_p & =  u_n + u_a  \\
& = \bar u_{MPC} \left( {{\bf \bar x}_0} \right) + {\bf K}\left( {{\bf x}_0 - {\bf \bar x}_0} \right) + \bar u_{MPC} \left( {\bf x}_0 \right).
\end{split}
\end{equation}
As presented in Sec. V-A, the actual state trajectory can be made to converge to the target points more effectively by using the joint scheme presented in (12) rather than  using $u_n$ or $u_a$ alone. The control laws $u_a$ (11) and $u_d$ (7) differ in terms of the constraints used in MPC (tightened and original constraints, respectively).

In practice, the tightened constraints used in tube-MPC, namely $\mathbb{\bar X}$, $\mathbb{\bar T}$, and $\mathbb{\bar U}$, can be estimated offline by using predefined disturbance bounds. However, tube-MPC methods with fixed uncertainty bounds often produce excessively conservative solution trajectories because worst-case disturbances  do not always occur during maneuvers. To reduce conservatism and improve system performance, $\mathbb{\bar X}$, $\mathbb{\bar T}$, and $\mathbb{\bar U}$ can be adaptively estimated online by detecting the variable disturbance ${\bf w}$. For example, the road curvature $\kappa$ is assumed to be detectable online in this study. Thus,  ${\bf w}$ and the corresponding tightened constraints (denoted as $\mathbb{\bar X}_{det}$/$\mathbb{\bar T}_{det}$/$\mathbb{\bar U}_{det}$)  can then be estimated.

\subsection{Interpolation Tube-MPC scheme}
Online estimations of $\mathbb{\bar X}_{det}$, $\mathbb{\bar T}_{det}$, and $\mathbb{\bar U}_{det}$ are less conservative than the offline constraints $\mathbb{\bar X}$, $\mathbb{\bar T}$, and $\mathbb{\bar U}$. However, $\mathbb{\bar X}_{det}$, $\mathbb{\bar T}_{det}$, and $\mathbb{\bar U}_{det}$ fluctuate, potentially resulting in unstable control inputs and system instability. To overcome this problem, interpolation \cite{Kog20} is applied to smooth the estimated tightened constraints.

Assume that the current detected road curvature is $\kappa _{det}$ and that the corresponding tightened constraints are $\mathbb{\bar X}_{det}$ := \{$\left. {\bar x}_{i \ne N} \right|\left| {\bar x}_{i \ne N} \right| \le \bar x_{det}$\}, $\mathbb{\bar T}_{det}$ := \{$\left. {\bar x}_{N} \right|\left| {\bar x}_{N} \right| \le \bar t_{det}$\}, and $\mathbb{\bar U}_{det}$ := \{$\left. {\bar u}_{i \ne N} \right|\left| {\bar u}_{i \ne N} \right| \le \bar u_{det}$\}. The original tightened constraints ($\mathbb{\bar X}_{det}$/$\mathbb{\bar T}_{det}$/$\mathbb{\bar U}_{det}$) can be improved using interpolation techniques incorporating looser ($\mathbb{\bar X}_{b}$/$\mathbb{\bar T}_{b}$/$\mathbb{\bar U}_{b}$) and tighter ($\mathbb{\bar X}_{s}$/$\mathbb{\bar T}_{s}$/$\mathbb{\bar U}_{s}$) constraints. The related constraint bounds are designed to satisfy the following assumption:
\begin{assumption}
\begin{subequations}
\begin{equation}
0 \le \bar x_{s}  < \bar x_{det}  < \bar x_{b} ,
\end{equation}
\begin{equation}
0 \le \bar t_{s}  < \bar t_{det}  < \bar t_{b} ,
\end{equation}
\begin{equation}
0 \le  \bar u_{s}  < \bar u_{det}  < \bar u_{b} ,
\end{equation}
\begin{equation}
{\bar x_{det}  - \bar x_{s} } = {\bar x_{b}  - \bar x_{det} } \equiv D_x \bar x_{\det } ,
\end{equation}
\begin{equation}
{\bar t_{det}  - \bar t_{s} } = {\bar t_{b}  - \bar t_{det} } \equiv D_t \bar t_{\det } ,
\end{equation}
\begin{equation}
{\bar u_{det}  - \bar u_{s} } = {\bar u_{b}  - \bar u_{det} } \equiv D_u \bar u_{\det } ,
\end{equation}
\end{subequations}
where $\bar x_{b}$, $\bar t_{b}$, and $\bar u_{b}$ cannot exceed the actual system constraint bounds. The following assumption is made in this paper: $D_ x $ = $D_ t $ = $D_ u $ $\equiv$ D.
\end{assumption}
To  fuse the aforementioned constraints, an interpolation variable sequence ${\bf L } \equiv \left\{ {{\bf \Lambda }_0 ,{\bf \Lambda }_1 ,...,{\bf \Lambda }_N } \right\}$ and ${\bf \Lambda } \equiv \left[ {\begin{array}{*{20}c}
   {\lambda _s } & {\lambda _{d} } & {\lambda _b }  \\
\end{array}} \right]^T $ are introduced into the optimization problem. Assumption 8 is an imposed design constraint that allows the optimizer to find optimal solutions more easily. The improved  tightened constraints $\mathbb{\bar X}_{int}$, $\mathbb{\bar T}_{int}$, and $\mathbb{\bar U}_{int}$ are then formulated by employing the following convex combinations:
\begin{assumption}
\begin{subequations}
\begin{equation}
{ \mathbb {\bar X}}_{int} = \lambda _s {\mathbb {\bar X}}_s  \oplus \lambda _{d} {\mathbb {\bar X}}_{det}  \oplus \lambda _b {\mathbb {\bar X}}_b ,
\end{equation}
\begin{equation}
{ \mathbb {\bar T}}_{int} = \lambda _s {\mathbb {\bar T}}_s  \oplus \lambda _{d} {\mathbb {\bar T}}_{det}  \oplus \lambda _b {\mathbb {\bar T}}_b ,
\end{equation}
\begin{equation}
{\mathbb {\bar U}}_{int} = \lambda _s {\mathbb {\bar U}}_s  \oplus \lambda _{d} {\mathbb {\bar U}}_{det}  \oplus \lambda _b {\mathbb {\bar U}}_b ,
\end{equation}
\end{subequations}
where the sets $\mathbb{\bar X}_{b}$/$\mathbb{\bar T}_{b}$/$\mathbb{\bar U}_{b}$ and $\mathbb{\bar X}_{s}$/$\mathbb{\bar T}_{s}$/$\mathbb{\bar U}_{s}$ are generated by scaling  $\mathbb{\bar X}_{det}$/$\mathbb{\bar T}_{det}$/$\mathbb{\bar U}_{det}$ with $D$ defined in (13).  
\end{assumption}
The MPC problem subject to the aforementioned tightened interpolation constraints (14) is called the interpolation tube-MPC problem (Problem 3), which is expressed as follows:

\textit{Problem 3}:
\begin{subequations}
\begin{equation}
\mathop {\min }\limits_{{\bf \bar X},{\bf \bar U},{\bf L}} \bar {J}_{int} = \bar{J}_s  + \bar{J}_t  + J_l ,
\end{equation}
\begin{equation}
\bar{J}_s  = \sum\limits_{i = 0}^{N - 1} {{\bf \bar x}_i^T {\bf Q \bar x}_i  + {\bf \bar u}_i^T {\bf R \bar u}_i } ,
\end{equation}
\begin{equation}
\bar{J}_t  = {\bf \bar x}_N^T {\bf P \bar x}_N ,
\end{equation}
\begin{equation}
 J_l  = \sum\limits_{i = 0}^N {{\bf \Lambda }_i^T {\bf S\Lambda }_i } 
\end{equation}
\end{subequations}
subject to
\begin{subequations}
\begin{equation}
{\bf \bar x}_{i + 1}  = {\bf \bar f}\left( {{\bf \bar x}_i ,{\bf \bar u}_i } \right) ,\quad 0 \le i < N,
\end{equation}
\begin{equation}
{\bf \bar x}_{i\ne N} \in \mathbb{\bar X}_{int}, 
\end{equation}
\begin{equation}
{\bf \bar x}_{N} \in \mathbb{\bar T}_{int}, 
\end{equation}
\begin{equation}
{\bf \bar u}_{i\ne N} \in \mathbb{\bar U}_{int},
\end{equation}
\begin{equation}
\lambda _{s}, \lambda _{d}, \lambda _{b}  > 0,
\end{equation}
\begin{equation}
\lambda _s  + \lambda _{d}  + \lambda _b  = 1.0,
\end{equation}
\end{subequations}
where the quadratic term $J_l$ (15d) with the weighting matrix ${\bf S}$ is leveraged to penalize deviations from 0 for the interpolation variables. The term $J_l$ (15d) and the constraints presented in (16e) and (16f) drive the interpolation variables to small positive numbers. To improve real-time controller performance, the tightened constraint calculation in (14) can be performed offline, and the results can be stored in a look-up table for online use.

The closed-loop properties of Problem 3 are described as follows.
\begin{proposition}
If Assumptions 5\textendash9 hold, Problem 3 is still convex because the operations presented in (14) for $\mathbb{\bar X}_{int}$/$\mathbb{\bar T}_{int}$/$\mathbb{\bar U}_{int}$ preserve the convexity of $\mathbb{\bar X}_{det} \subseteq \mathbb{\bar X}$, $\mathbb{\bar T}_{det} \subseteq \mathbb{\bar T}$, and $\mathbb{\bar U}_{det} \subseteq \mathbb{\bar U}$. Therefore, Problem 3 has the same closed-loop properties as does the standard tube-MPC scheme (Proposition 2).
\end{proposition}
A major discrepancy between the proposed interpolation scheme and the standard tube-MPC scheme is that the proposed scheme includes more degrees of freedom, enabling it to find a better solution with an acceptable increase in computational complexity. Moreover, in contrast to the interpolation tube-MPC scheme presented in \cite{Kog20}, in which different control gains are used to construct multiple tubes for blending, a scaling parameter ($D$) is employed for this purpose in the proposed scheme [(13) and (14)], which reduces the tuning difficulty.

\begin{figure*}[!t]
{\includegraphics*[width=0.25\linewidth]{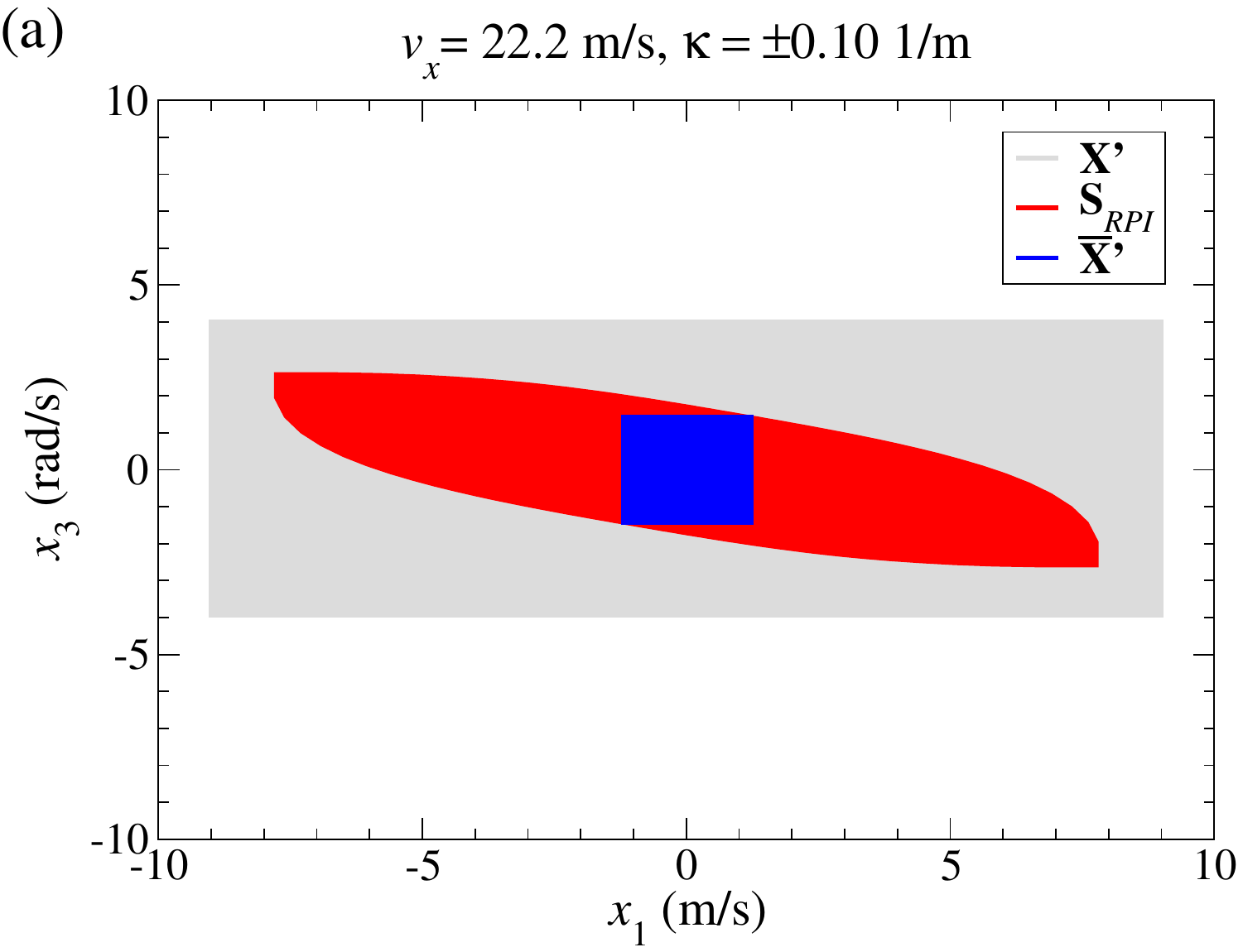}}%
{\includegraphics*[width=0.25\linewidth]{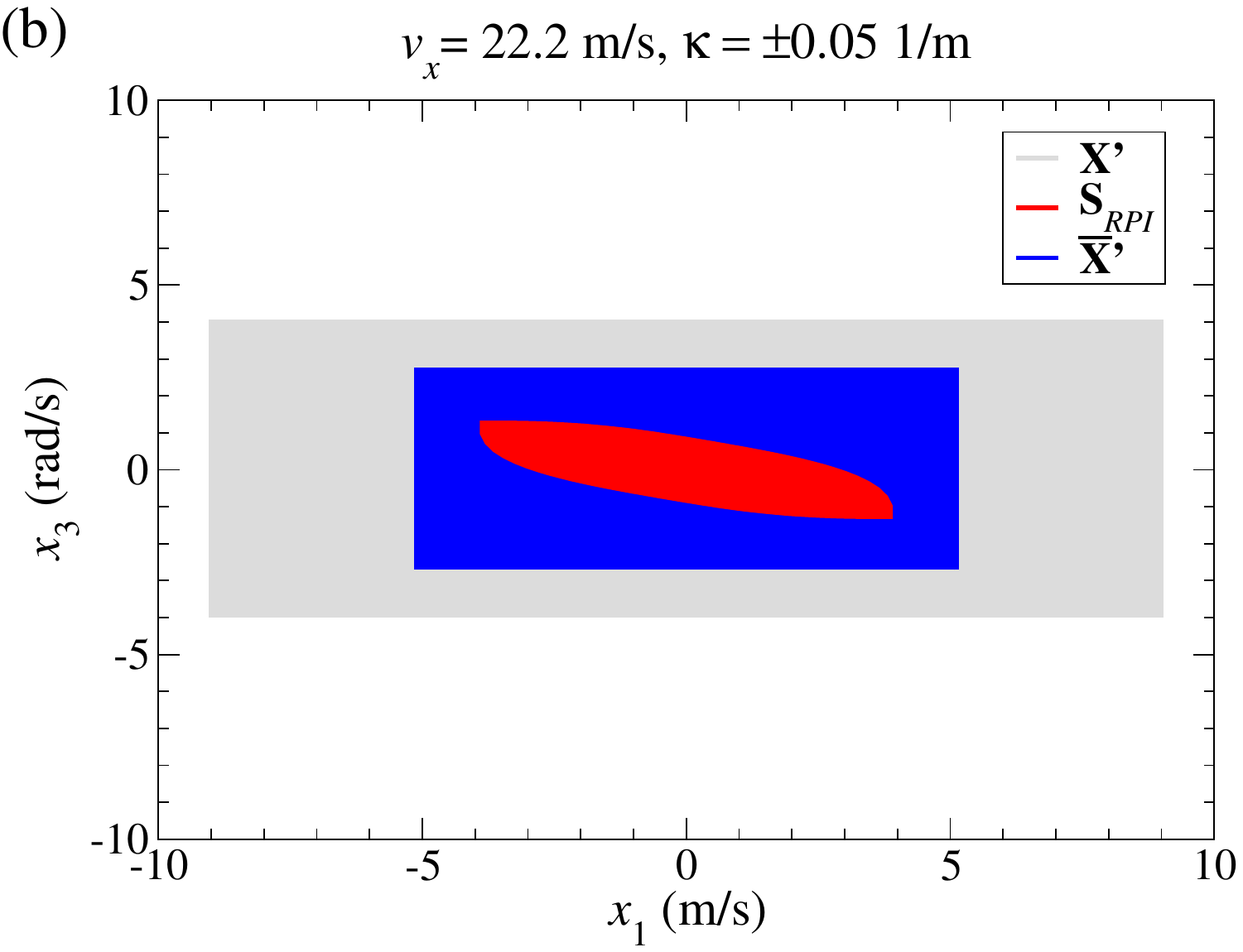}}%
{\includegraphics*[width=0.25\linewidth]{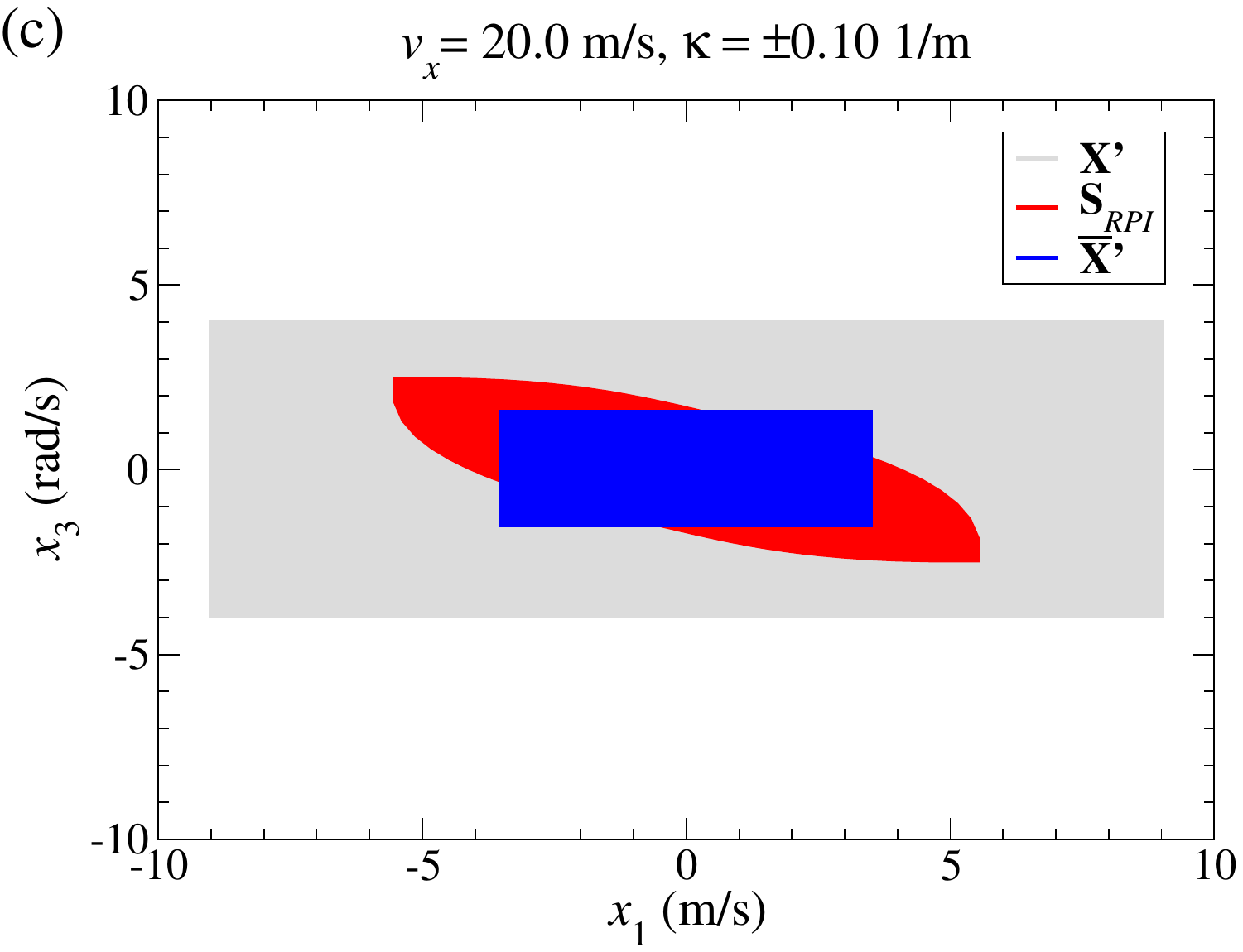}}%
{\includegraphics*[width=0.25\linewidth]{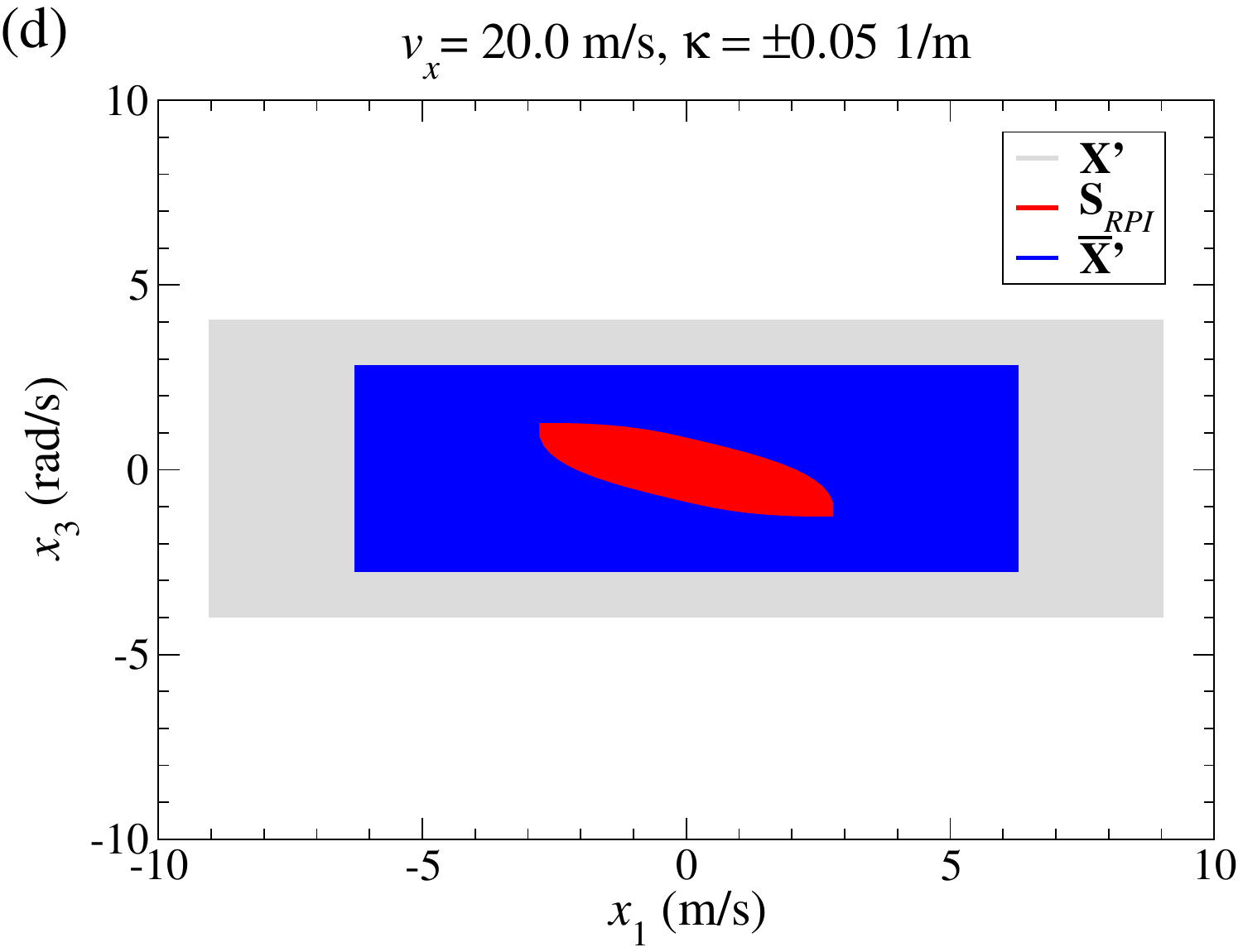}}%
\caption{Visualization of sets $\mathbb{X}'$, $\mathbb{\bar X}'$, and $\mathbb{S}_{RPI}$ when $v_x$ = 20.0 or 22.2 m/s and $\kappa$ = $\pm$0.05 and $\pm$0.1 1/m. Here, $x_1  \equiv \dot \Delta $ and $x_3  \equiv \dot \theta $.}
\end{figure*}

\begin{figure}[t]
{\includegraphics*[width=0.51\linewidth]{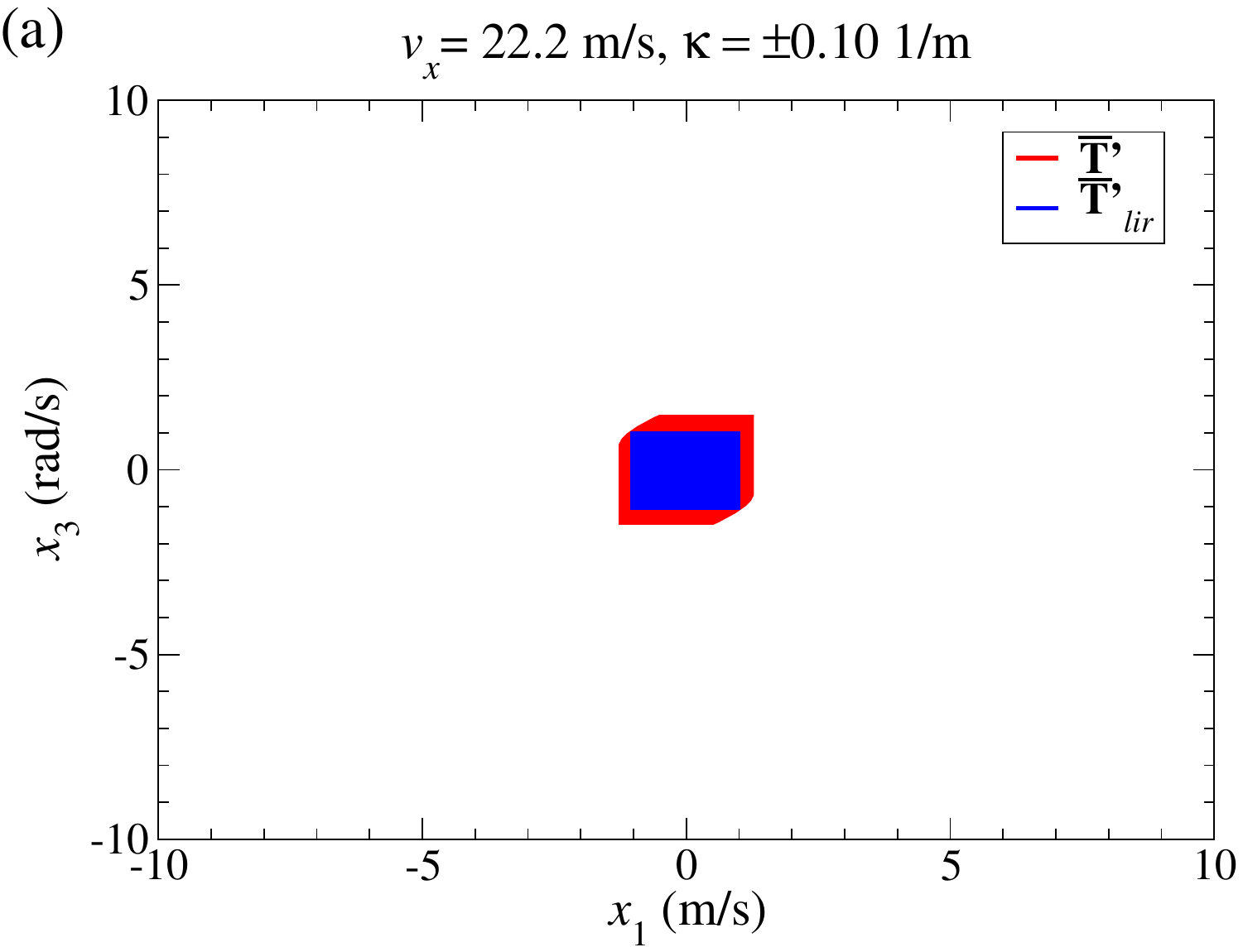}}%
{\includegraphics*[width=0.51\linewidth]{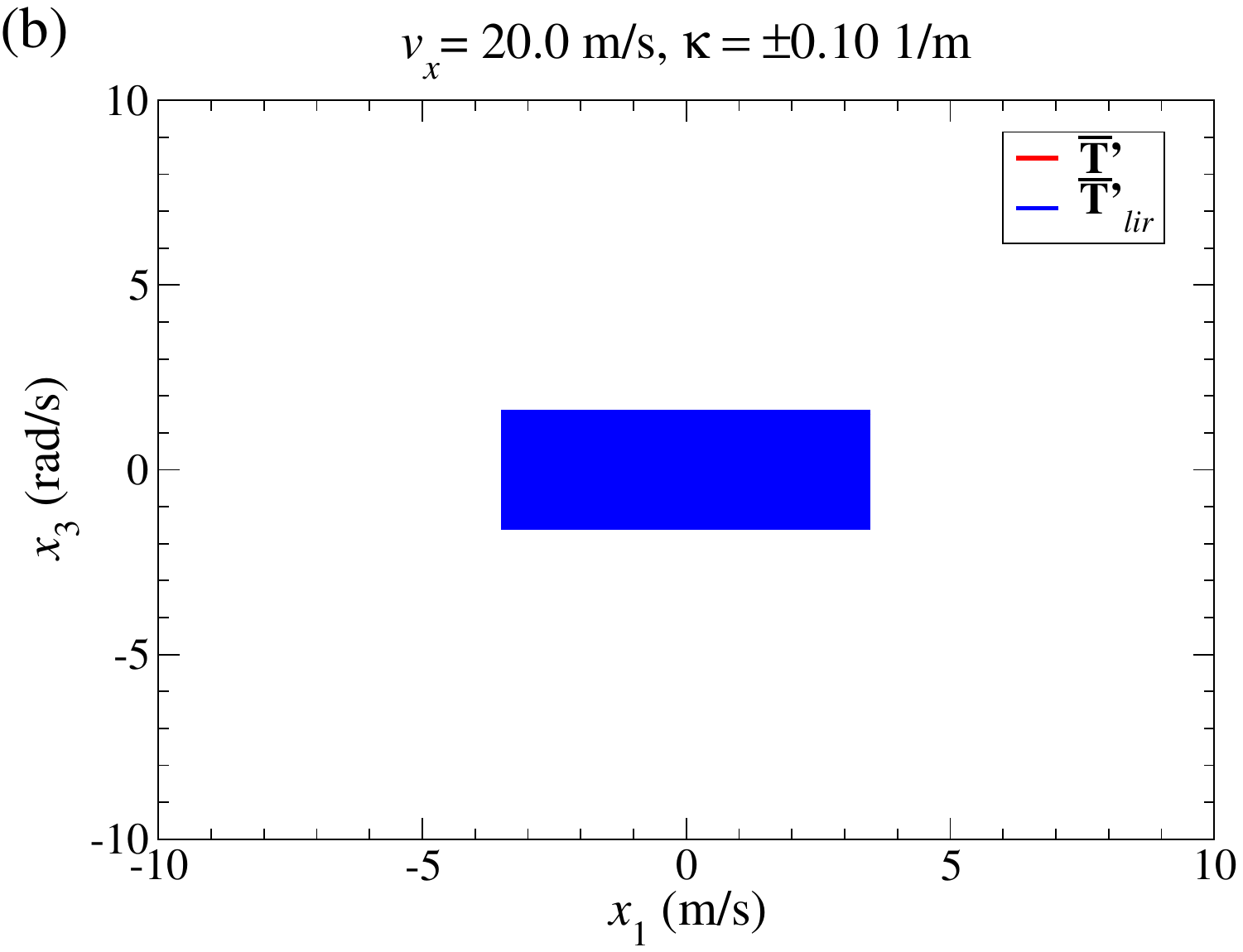}}%
\caption{Visualization of sets $\mathbb{\bar T}'$ and $\mathbb{\bar T}'_{lir}$ when $v_x$ = 20.0 or 22.2 m/s and $\kappa$ = $\pm$0.1 1/m. In (a), $\mathbb{\bar T}'$ is defined by 14 inequalities, and the corresponding vertex coordinates in the first and third quadrants are identical to those of $\mathbb{\bar X}'$ in Fig. 1(a).}
\end{figure}

\subsection{Computation of Tightened Constraints}
For tube-based approaches, the calculation of tightened constraints requires the use of set algebra; however, this approach is computationally expensive even for a four-dimensional system \cite{Kha21} because the calculation involves many polytope vertices. A solution to this problem is to project the original high-dimensional system onto low-dimensional subsystems \cite{Kha21, Wan22}. In the discretized lane-keeping  model presented in (1), the disturbance term ${\bf w}$ is dominated by two state variables, $\dot \Delta$ and $\dot \theta$. Thus, the computation of the tightened constraints for the original four-dimensional system ${\bf x}$ is approximated by considering the two-dimensional subsystem ${\bf x}' \equiv  \left[ {\begin{array}{*{20}c} {\dot \Delta }  & {\dot \theta }  \\
\end{array}} \right]^T \in \mathbb{X}'/\mathbb{T}'$ with the control vector ${\bf u}' \equiv  \left[  \delta  \right] \in \mathbb{U}'$ and the disturbance ${\bf w}' \in \mathbb{W}'$. The relevant discrete-time dynamics model is represented as follows \cite{Lev22}:
\begin{equation}
{\bf x'}_{i + 1}  = {\bf A'x'}_i  + {\bf B'u'}_i   + {\bf w'}.
\end{equation}
The corresponding model matrices are expressed as follows:
\[
\begin{array}{l}
 {\bf A'} = \left[ {\begin{array}{*{20}c}
   {a _{22} } & {a _{24}  - v_x dt}  \\
   {a _{42} } & {a _{44} }  \\
\end{array}} \right], \\ 
 {\bf B'} = \left[ {\begin{array}{*{20}c}
   {b _1 }  \\
   {b _2 }  \\
\end{array}} \right], \quad {\bf w'} = \kappa\left[ {\begin{array}{*{20}c}
   {c _1 }  \\
   {c _2 }  \\
\end{array}} \right]. \\ 
 \end{array}
\]
The LQR gain ${\bf K'} \equiv \left[ {\begin{array}{*{20}c}
   {K'_0 } & {K'_1 }  \\
\end{array}} \right]$ for the aforementioned subsystem is computed using the parameters listed in Table I.

The tightened constraints are computed using an outer-bounding tube of the disturbance invariant set \cite{Raw17}. Because ${\bf A}_{K'}'  \equiv {\bf A}' + {\bf B}'{\bf K}'$ is stable,  the uncertainty sets can be defined as follows:
\begin{equation}
\mathbb{S}_i \equiv \sum\limits_{j = 0}^{i - 1} {{\bf A}_{K'}'^j \mathbb W'} .
\end{equation}
There exists a finite integer $N$ and a scalar $\alpha  \in \left[ {0,1)} \right.$ such that ${\bf A}_{K'}'^N \mathbb W' \subseteq  \alpha \mathbb W'$ and ${\bf K}'{\bf A}_{K'}'^N \mathbb W' \subseteq  \alpha {\bf K}'\mathbb W'$. It follows that
\begin{equation}
\mathbb{S}_\infty   \subseteq \frac{1}{{1 - \alpha }}\mathbb{S}_N  ,
\end{equation}
where ${{\mathbb{S}_N } \mathord{\left/
 {\vphantom {{S_N } {1 - \alpha }}} \right.
 \kern-\nulldelimiterspace} {1 - \alpha }} $ is a robust positively invariant (RPI) set, denoted as $\mathbb{S}_{ RPI}$. Because $\mathbb W'$ is box constrained, it is sufficient to check only its vertices ${\bf w}'_v  \equiv \left[ {\begin{array}{*{20}c}
   \pm {w_{v0} } & \pm {w_{v1} }  \\
\end{array}} \right]^T$ to compute $\alpha$ using the following formula \cite{Raw17}:
\begin{equation}
\alpha  = \max \left( {\frac{{\left\| {{\bf A'}_{K'}^N {\bf w'}_v } \right\|_\infty  }}{{\left\| {{\bf w'}_v } \right\|_\infty  }},\frac{{\left\| {{\bf K'A'}_{K'}^N {\bf w'}_v } \right\|_\infty  }}{{\left\| {{\bf K'w'}_v } \right\|_\infty  }}} \right).
\end{equation}
Alternatively, $\alpha$ can be derived by solving the following linear programming problem:
\begin{equation}
\mathop {\max }\limits_{i = 0}^{n_r-1 } \left( {\mathop {\max }\limits_{\bf z} \alpha _i  = {\bf F}_i {\bf A'}_{K'}^N {\bf z}} \right)
\end{equation}
subject to
\begin{equation}
{\bf Fz} \le {\bf 1},
\end{equation}
where ${\bf z} \equiv \left[ {\begin{array}{*{20}c}
   {z_0 } & {z_1 }  \\
\end{array}} \right]^T $ and
\[
 {\bf F} \equiv \left( {\begin{array}{*{20}c}
   {\frac{{ - 1}}{{w_{v0} }}} & 0  \\
   {\frac{1}{{w_{v0} }}} & 0  \\
   0 & {\frac{{ - 1}}{{w_{v1} }}}  \\
   0 & {\frac{1}{{w_{v1} }}}  \\
   {  \frac{{-K'_0 }}{{\left( {K'_0 w_{v0}  + K'_1 w_{v1} } \right)}}} & {  \frac{{-K'_1 }}{{\left( {K'_0 w_{v0}  + K'_1 w_{v1} } \right)}}}  \\
   {\frac{{K'_0 }}{{\left( {K'_0 w_{v0}  + K'_1 w_{v1} } \right)}}} & {\frac{{K'_1 }}{{\left( {K'_0 w_{v0}  + K'_1 w_{v1} } \right)}}}  \\
\end{array}} \right).
\]
The values of $n_r$ in (21) are the row numbers of the matrix $\bf F$. Consequently, the tightened stage constraints can be computed as follows:
\begin{subequations}
\begin{equation}
\mathbb{\bar X}' = \mathbb{X}' \ominus \mathbb{S}_{RPI} ,
\end{equation}
\begin{equation}
\mathbb{\bar U}' = \mathbb{U}' \ominus {\bf K}' \mathbb{S}_{RPI} .
\end{equation}
\end{subequations}
These constraints can be expressed as a set of linear inequalities as follows:
\begin{equation}
{\bf F'x'} + {\bf G'u'} \le {\bf h'}.
\end{equation}
Subsequently, the maximal positively invariant set \cite{Kou16} under the  tightened stage constraints (24) is selected as the tightened terminal constraint $\mathbb{\bar T}'$, which can be represented as follows: 
\begin{equation}
\mathbb{\bar T}' = \left\{ {\left. {{\bf x'}} \right|\left( {{\bf F'} + {\bf G'K'}} \right)\left( {{\bf A'} + {\bf B'K'}} \right)^i {\bf x'} \le {\bf h'},i = 0,...,N_\nu  } \right\}.
\end{equation}
The term $N_\nu$ denotes the smallest positive integer that can be computed by solving  the following linear programming problem:
\begin{equation}
\mathop {\max }\limits_{{\bf x'}} \left( {{\bf F'} + {\bf G'K'}} \right)_j \left( {{\bf A'} + {\bf B'K'}} \right)^{n + 1} {\bf x'},\quad j = 1,...,n_c 
\end{equation}
subject to
\begin{equation}
\left( {{\bf F'} + {\bf G'K'}} \right)\left( {{\bf A'} + {\bf B'K'}} \right)^i {\bf x'} \le {\bf h'},\quad i = 0,...,n,
\end{equation}
where $n = 1,...,N_\nu$ and $n_c$ indicates the number of rows in the matrix ${{\bf F'} + {\bf G'K'}}$.

To enable real-time applications, $\mathbb{\bar X}'$/$\mathbb{\bar U}'$/$\mathbb{\bar T}'$ can be precomputed offline, and the results can be stored in a lookup table for online applications \cite{Lu25}. This table is termed the $\kappa$-table throughout this paper. It indicates corresponding tightened constraints when $\mathbb{ X}' $/$\mathbb{ U}'$, $v_x$, and $\kappa$ are provided. Specifically, given $\mathbb{ X}' $/$\mathbb{ U}'$ and $v_x$, the tightened constraints $\mathbb{\bar X}' $/$\mathbb{\bar U}'$/$\mathbb{\bar T}'$ corresponding to 201 uniformly distributed data points for $\kappa$ over the range of [-0.1, 0.1] 1/m are provided in the aforementioned table. The tightened constraints required for planning and control are derived online by looking up the nearest-neighbor state of the current input in the table. Fig. 1 presents typical results for computations for the constraint region of actual system, nominal system, and disturbance invariant set for $v_x$ = 20.0 or 22.2 m/s and $\kappa $ = $\pm$0.05 or $\pm$0.1 1/m. The size of the nominal constraint region $\mathbb{\bar X}'$ decreases when that of the $\mathbb{S}_{RPI}$ region increases. Consequently, the smallest set of feasible nominal states is obtained when $v_x$ = 22.2 m/s and $\kappa $ = $\pm$0.1 1/m [Fig. 1(a)]. Fig. 2(a) and 2(b) displays results obtained in terminal constraint ($\mathbb{\bar T}'$) computations under the stage constraints ($\mathbb{\bar X}'$) illustrated in Fig. 1(a) and 1(c), respectively. In particular, the number of inequalities defining $\mathbb{\bar T}'$ may be large, which can result in a heavy online computational burden. To reduce this burden, the largest interior rectangular constraint region ($\mathbb{\bar T}'_{lir}$) is used in the online optimization (Fig. 2). Under a small disturbance [e.g., that shown in Fig. 1(c)], $\mathbb{\bar T}' = \mathbb{\bar T}'_{lir} = \mathbb{\bar X}'$ [Fig. 2(b)].

The aforementioned method, which involves mapping the currently detected $\kappa$ value to specified tightened constraints ($\mathbb{\bar X}' $/$\mathbb{\bar T}' $/$\mathbb{\bar U}'$), has two major drawbacks. First, the input state does not precisely match the nearest-neighbor state in the look-up table. Second, fluctuations caused in the detected $\kappa$ value by unknown external noise can result in constraint fluctuations, potentially leading to instability of the controlled system. As described in Sec. III-C, these problems are mitigated in the present study by using the proposed interpolation algorithm to compute improved tightened constraints.

\section{itube-CILQR Algorithm}
This section describes the proposed itube-CILQR algorithm. Convex and smooth exponential barrier functions are used to convert Problem 3 into an equivalent unconstrained  problem for the CILQR algorithm \cite{Che17,Liu24}. As displayed in Fig. 3, similar to typical quadratic cost terms, the exponential barrier function reaches its minimum value at the origin, which preserves the convexity of the objective functions. This property is crucial for barrier-function-based MPC algorithms \cite{Wil04}.

The explicit form of Problem 4 is expressed as follows:

\textit{Problem 4}:
\begin{subequations}
\begin{equation}
\mathop {\min }\limits_{{\bf \bar X},{\bf \bar U},{\bf L}} \bar {J}_{int}' = \bar{J}_s  + \bar{J}_t  + J_l + J_{x0}+ J_{x1}+ J_{u}+ J_{lu}+ J_{ls},
\end{equation}
\begin{equation}
\bar{J}_s  = \sum\limits_{i = 0}^{N - 1} {{\bf \bar x}_i^T {\bf Q \bar x}_i  + {\bf \bar u}_i^T {\bf R \bar u}_i } ,
\end{equation}
\begin{equation}
\bar{J}_t  = {\bf \bar x}_N^T {\bf P \bar x}_N ,
\end{equation}
\begin{equation}
 J_l  = \sum\limits_{i = 0}^N {{\bf \Lambda }_i^T {\bf S\Lambda }_i } ,
\end{equation}
\begin{equation}
\begin{split}
&J_{x0}  = \\
&q_s \sum\limits_{k \in \left\{ {0,2} \right\}} {\sum\limits_{i = 0}^{N-1} {\exp \left( {\bar x_{k,\min }  - \bar x_{k,i} } \right) + \exp \left( {\bar x_{k,i}  - \bar x_{k,\max } } \right)} } \\
&+{\exp \left( {\bar t_{k,\min }  - \bar x_{k,N} } \right) + \exp \left( {\bar x_{k,N}  - \bar t_{k,\max } } \right)},
\end{split}
\end{equation}
\begin{equation}
\begin{split}
&J_{x1}  = \\
&q_s  \sum\limits_{k \in \left\{ {1,3} \right\}}  \sum\limits_{i = 0}^{N-1} { \exp  { \left[ { - \left( {\lambda _{s,i} \bar x_{k,s}  + \lambda _{d,i} \bar x_{k,det }  + \lambda _{b,i} \bar x_{k,b} } \right) - \bar x_{k,i} } \right]} }  \\
&+  \exp  { \left[ {\bar x_{k,i}  - \left( {\lambda _{s,i} \bar x_{k,s}  + \lambda _{d,i} \bar x_{k,det }  + \lambda _{b,i} \bar x_{k,b} } \right)} \right]} \\
&+  \exp  { \left[ { - \left( {\lambda _{s,N} \bar t_{k,s}  + \lambda _{d,N} \bar t_{k,det }  + \lambda _{b,N} \bar t_{k,b} } \right) - \bar x_{k,N} } \right]} \\
&+  \exp  { \left[ {\bar x_{k,N}  - \left( {\lambda _{s,N} \bar t_{k,s}  + \lambda _{d,N} \bar t_{k,det }  + \lambda _{b,N} \bar t_{k,b} } \right)} \right]} ,
\end{split}
\end{equation}
\begin{equation}
\begin{split}
&J_{u}  = q_{u}\sum\limits_{i = 0}^{N - 1} { \exp  { \left[ { - \left( {\lambda _{s,i} \bar \delta _s  + \lambda _{d,i} \bar \delta _{det}  + \lambda _{b,i} \bar \delta _b } \right) - \bar\delta _i } \right]} } \\
&+  \exp  { \left[ {\bar\delta _i  - \left( {\lambda _{s,i} \bar \delta _s  + \lambda _{d,i} \bar \delta _{det }  + \lambda _{b,i} \bar \delta _b } \right)} \right]} ,
\end{split}
\end{equation}
\begin{equation}
\begin{split}
J_{lu}  = q_{l1} \sum\limits_{k \in \left\{ {s,d,b} \right\}} {\sum\limits_{i = 0}^N {\exp \left( { - \lambda _{k,i} } \right) + \exp \left( {\lambda _{k,i}  - 1.0} \right)} } ,
\end{split}
\end{equation}
\begin{equation}
\begin{split}
&J_{ls}  = q_{l2} \sum\limits_{i = 0}^N {\exp \left\{ {q_{l2} \left[ {1.0 - \left( {\lambda _{s,i}  + \lambda _{d,i}  + \lambda _{b,i} } \right)} \right]} \right\}}  \\
&+ \exp \left\{ {q_{l2} \left[ {\left( {\lambda _{s,i}  + \lambda _{d,i}  + \lambda _{b,i} } \right) - 1.0} \right]} \right\}
\end{split}
\end{equation}
\end{subequations}
subject to
\begin{equation}
{\bf\bar x}_{i + 1}  = {\bf \bar f}\left( {{\bf \bar x}_i ,{\bf \bar u}_i } \right) ,\quad 0 \le i < N.
\end{equation}
Here, $\bar x_{1, det } $/$\bar t_{1, det } $, $\bar x_{3, det } $/$\bar t_{3, det } $, and $\bar \delta_{det } $ are the upper bounds of the tightened constraints for $\bar x_1$, $\bar x_3$, and $\bar\delta$, respectively. These upper bounds are derived through online lookup of the $\kappa$-table on the basis of the detected road curvature. The symbols $\bar x_{1, s } $/$\bar x_{1, b } $, $\bar t_{1, s } $/$\bar t_{1, b } $, $\bar x_{3, s } $/$\bar x_{3, b } $, $\bar t_{3, s } $/$\bar t_{3, b } $, and $\bar \delta_{s } $/$\bar \delta_{b } $ indicate the corresponding designed constraint bounds, which are described in Sec. III-C. The parameters $\bar x_{0/2, min } $ ($\bar t_{0/2, min } $) and $\bar x_{0/2, max } $ ($\bar t_{0/2, max } $) are the  bounds of the original constraints of $ x_{0/2}$. Moreover, the scalars $q_{s}$, $q_{u}$, and $q_{l1/2}$ denote penalty parameters. The penalty parameters  selected in this study are listed in Table II. In particular, ${\lambda _{d} }$ was set to a constant value of $1.0-2D$ (13) to reduce the number of variables and thus decrease the computational cost.

\begin{figure}[!t]
\centerline{\includegraphics[scale=0.215]{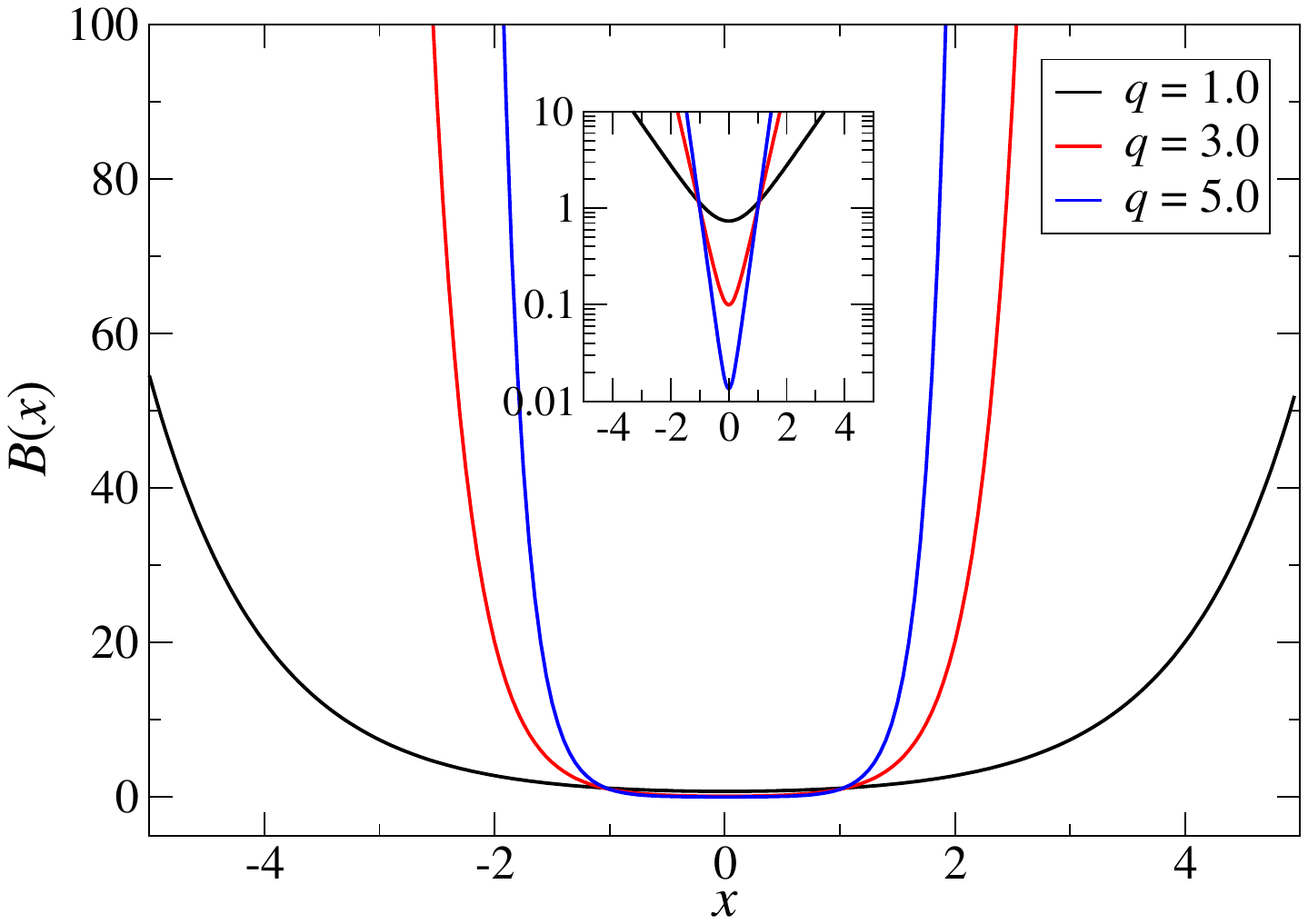}}
\caption{Profile of an example exponential barrier function $B\left( x \right) = \exp \left[ {q\left( {x - 1} \right)} \right] + \exp \left[ {q\left( { - 1 - x} \right)} \right]$, where $x$ can be a state or control variable. As the penalty parameter $q$ increases, $B(x)$ increases rapidly when $x > 1$ and $x < -1$ but approaches zero when $x =0$.}   
\end{figure}

To solve  Problem 4, the ILQR algorithm \cite{Tas14} and Newton's descent  method \cite{Lee23} are used to simultaneously optimize state, control, and interpolation variables iteratively through alternating backward and forward propagation steps. The relevant update laws for these variables are expressed as follows:
\begin{subequations}
\begin{equation}
{\bf \hat u}_i  = {\bf \bar u}_i  + {\bf \tilde k}_i  + {\bf \tilde K}_i \left( {{\bf \hat x}_i  - {\bf \bar x}_i } \right),
\end{equation}
\begin{equation}
{\bf \hat x}_{i + 1}  = {\bf \bar f}\left( {{\bf \hat x}_i ,{\bf \hat u}_i } \right),
\end{equation}
\begin{equation}
{\bf \hat \Lambda }_i  = {\bf \Lambda }_i  - \left[ {{\bf H}\bar J_{{\mathop{ int}} } '\left( {{\bf \Lambda }_i } \right)} \right]^{ - 1} \nabla \bar J_{{\mathop{ int}} } '\left( {{\bf \Lambda }_i } \right),
\end{equation}
\end{subequations}
where the control gains ${\bf \tilde K}_i $ and ${\bf \tilde k}_i $ are computed using the analytical formulas presented in \cite{Tas14} and the Hessian matrix ${\bf H}\bar J_{{\mathop{ int}}}'\left( {{\bf \Lambda}_i } \right) \succ 0$. 

The control scheme can be modified by varying the cost functions, input state  ${\bf \hat x}_0$, and control input $u$ to the actual system. The control schemes used in this study are described as follows. The itube-CILQR algorithm solves Problem 4 by using the control action $u$ from $u_p$ (12). The tube-CILQR-up algorithm solves Problem 2 by employing the CILQR and $u_p$ (12). The tube-CILQR-un algorithm solves Problem 2 by adopting the CILQR with the control action $u$ from $u_n$ (10). The tube-CILQR-ua algorithm solves Problem 2 by using the CILQR with the control action $u_a$ (11). Finally, the nominal CILQR algorithm solves Problem 1 by using the CILQR with the control law $u_d$ (7). In this study, Problems 1\textendash3 were also solved using five methods based on the interior-point optimizer (IPOPT) \cite{Get24, Wac06, Sha24} for comparison: itube-MPC, tube-MPC-up, tube-MPC-un, tube-MPC-ua, and nominal MPC.

\begin{figure*}[!t]
\centering
{\includegraphics*[width=0.33\linewidth]{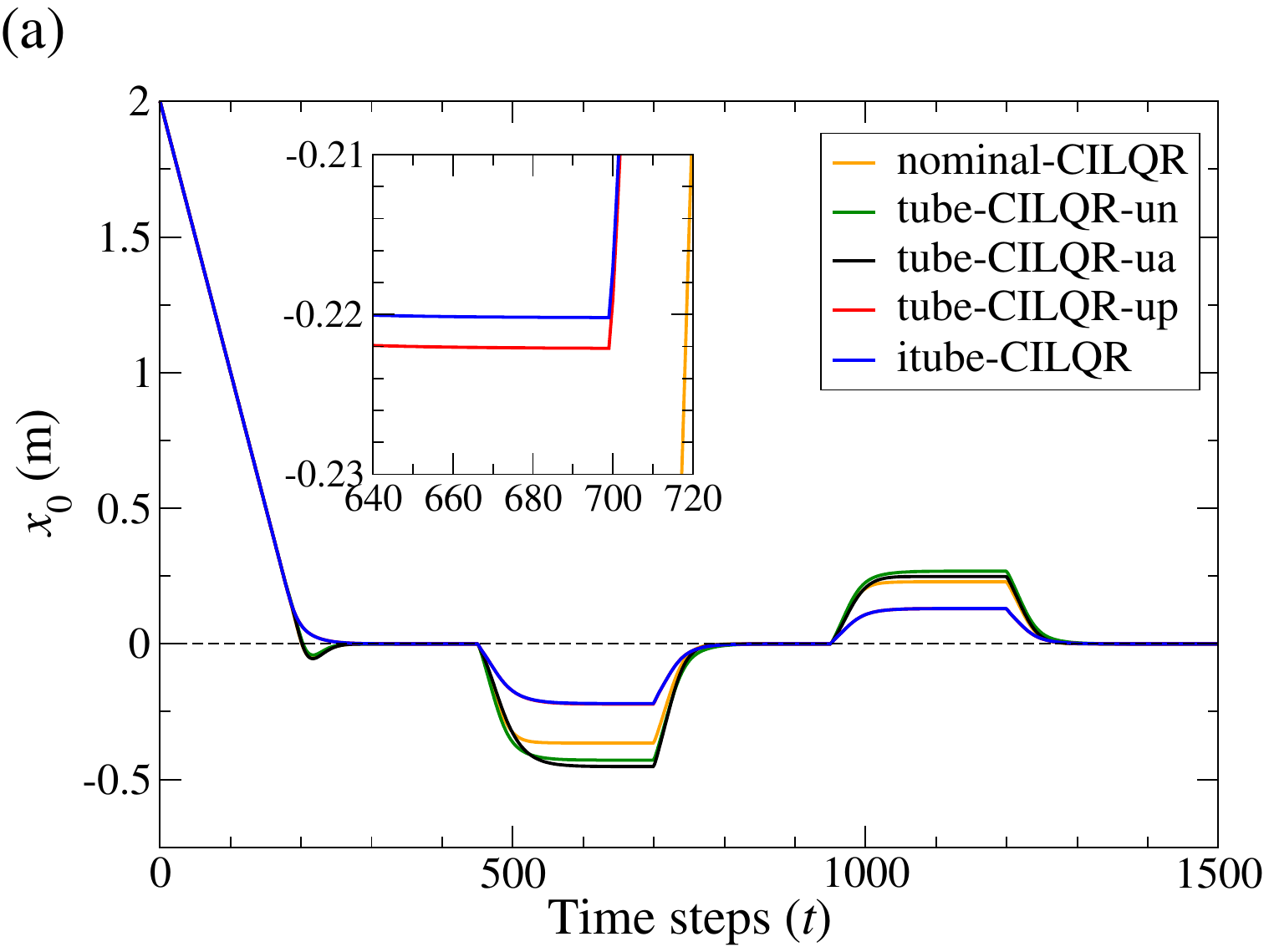}}%
{\includegraphics*[width=0.33\linewidth]{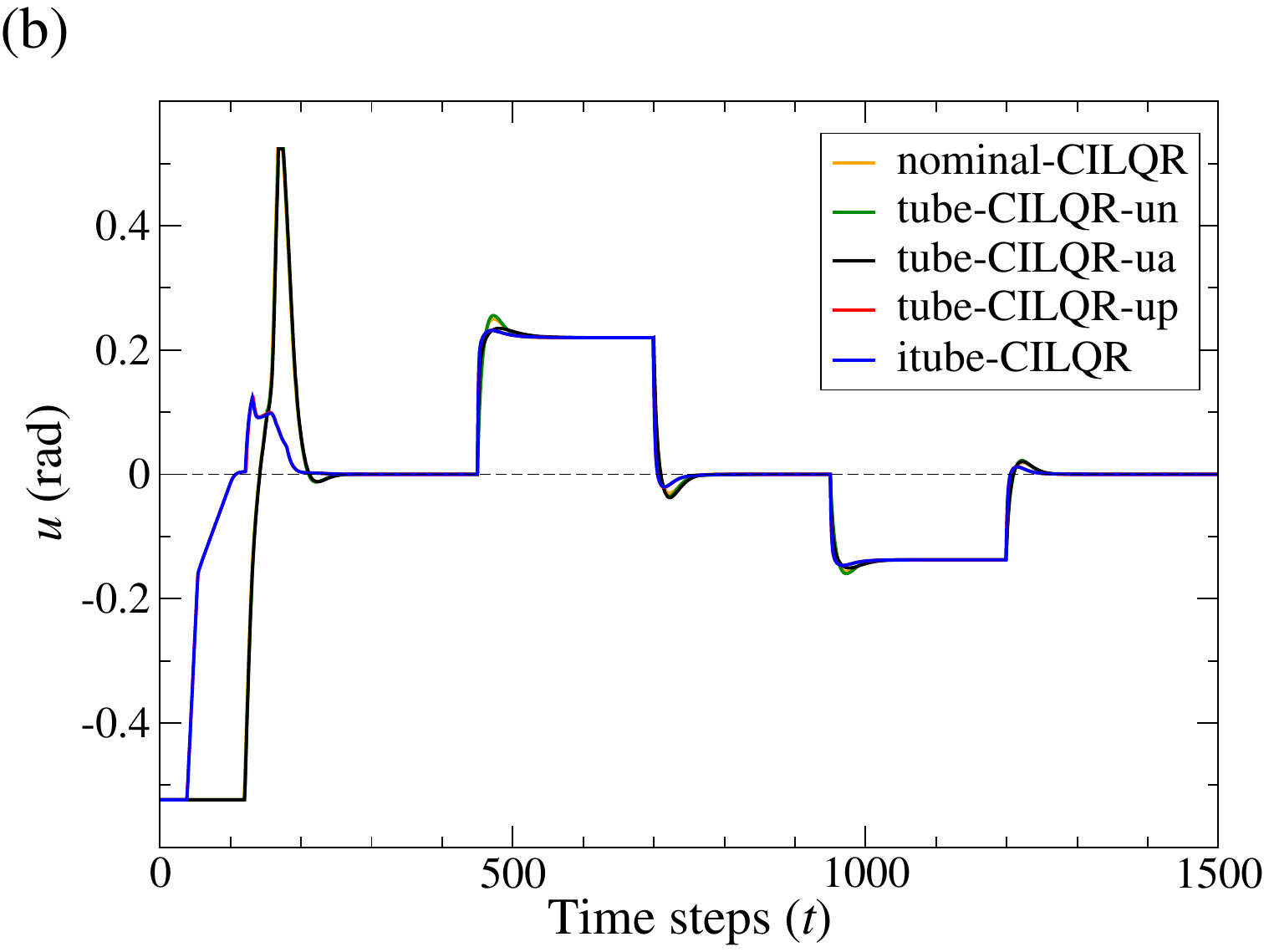}}%
{\includegraphics*[width=0.33\linewidth]{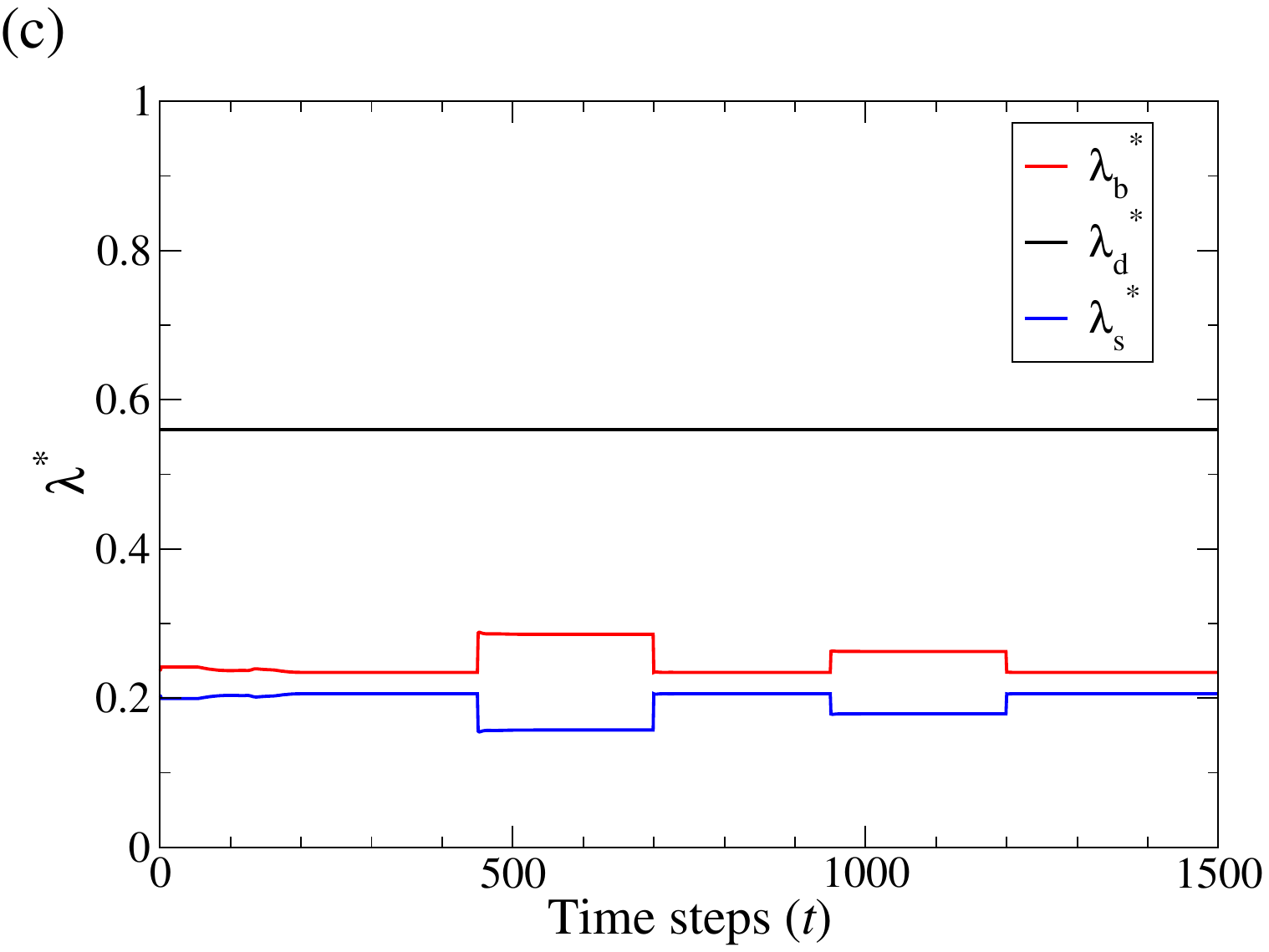}}%
\caption{Numerical simulation results obtained with different CILQR algorithms when $v_x$ = 20.0 m/s, $D$ = 0.22, $\left| {x_1 } \right| \le$ 9.0 m/s, and $\left| {x_3 } \right| \le$ 4.0 rad/s. (a) Trajectory of the actual system state $x_0$. (b) Trajectory of the actual system control input. (c) Trajectories of the first components of the optimal interpolation variable sequences for $x_0  \equiv  \Delta $ and $u \equiv \delta$. In (a), the $x_0$ values obtained with the tube-CILQR-up and itube-CILQR algorithms at $t$ = 700 are -0.2221 and  -0.2201 m, respectively. In (c), $\Delta \lambda  $ =  0.0331,  0.1283, and  0.0834 at $t$ = 0, 600, and 1100, respectively. The experimental results obtained at different $D$ values are presented in Table III.} 
\end{figure*}

\begin{figure*}[!t]
{\includegraphics*[width=0.25\linewidth]{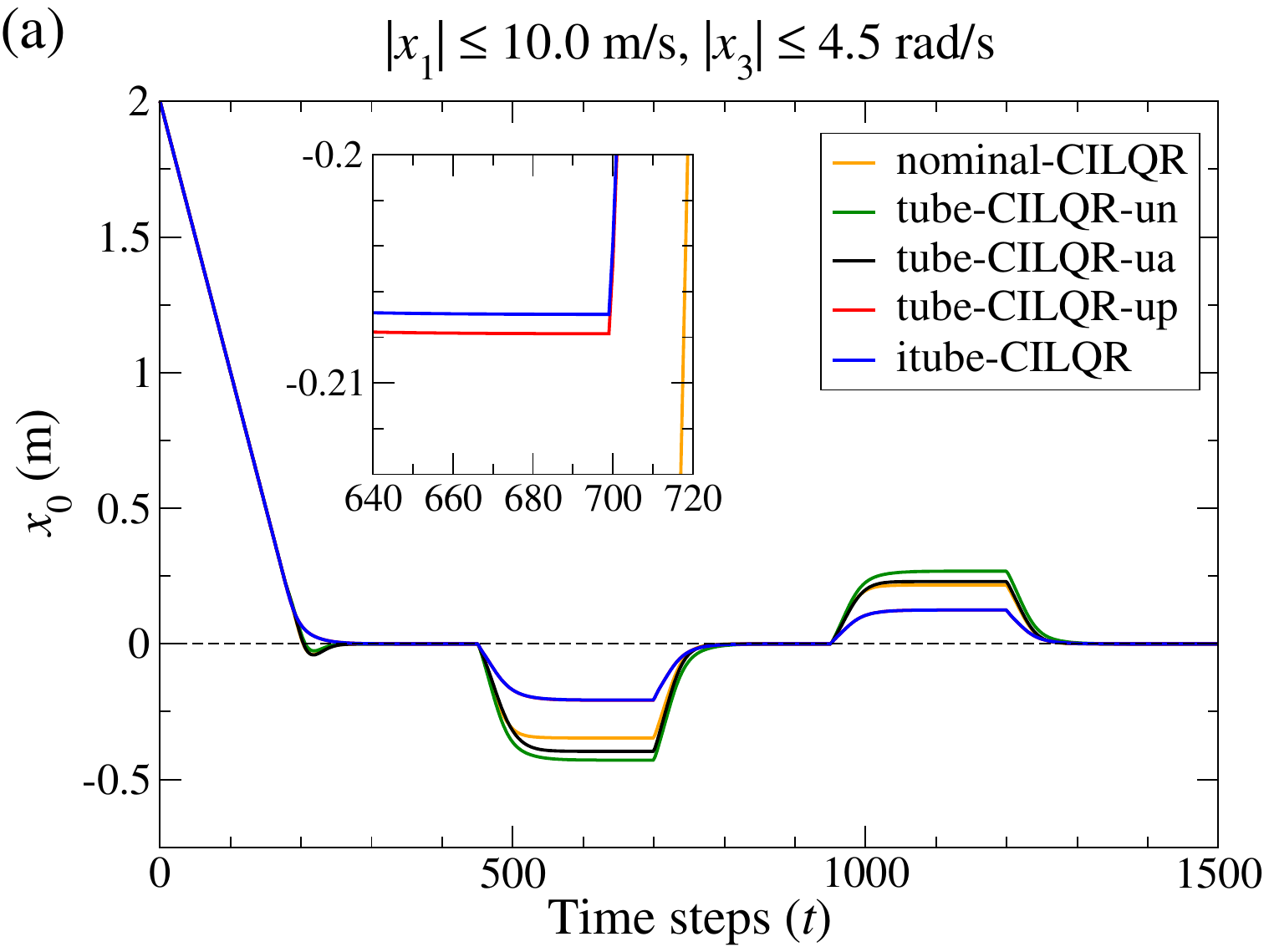}}%
{\includegraphics*[width=0.25\linewidth]{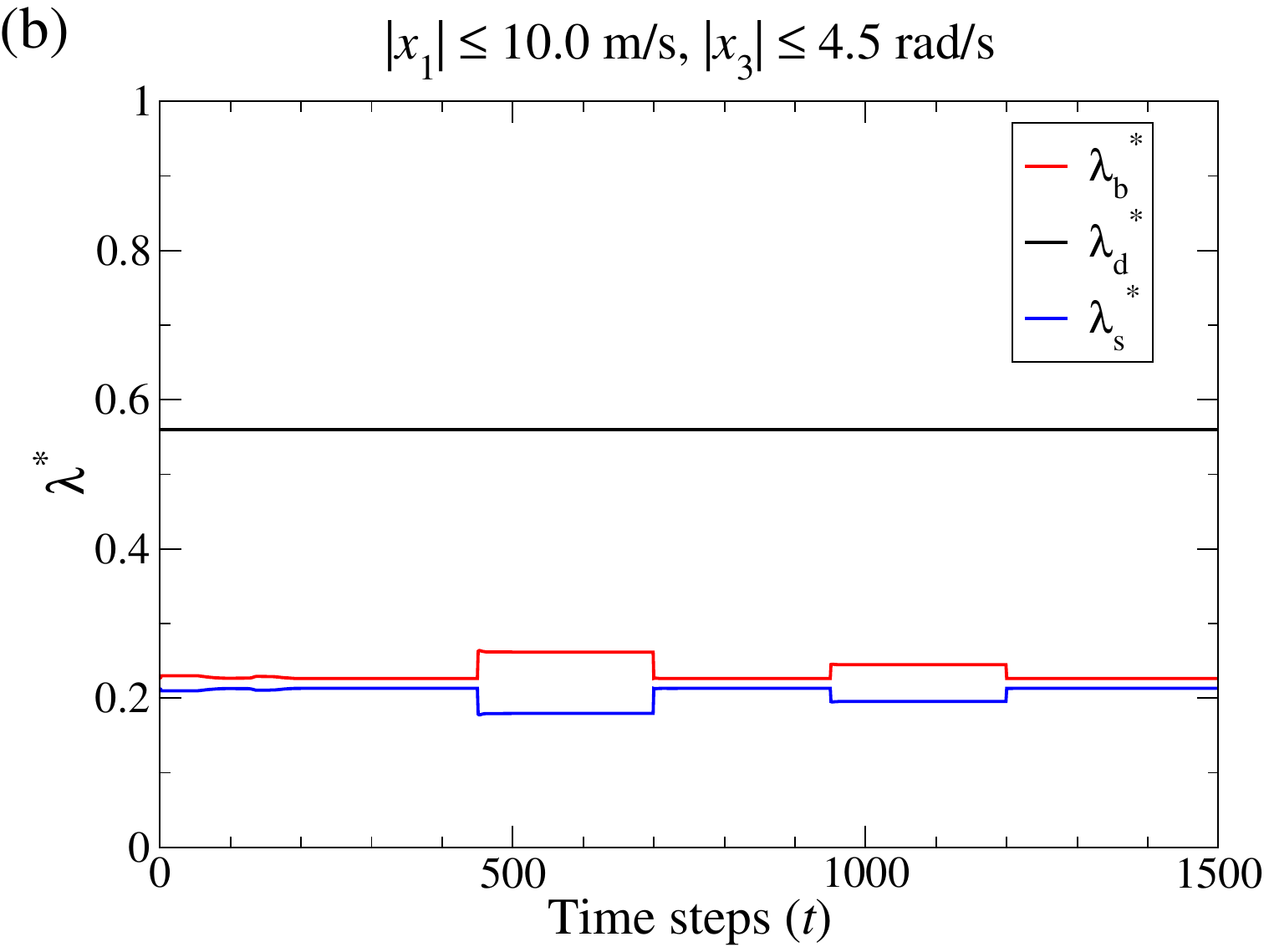}}%
{\includegraphics*[width=0.25\linewidth]{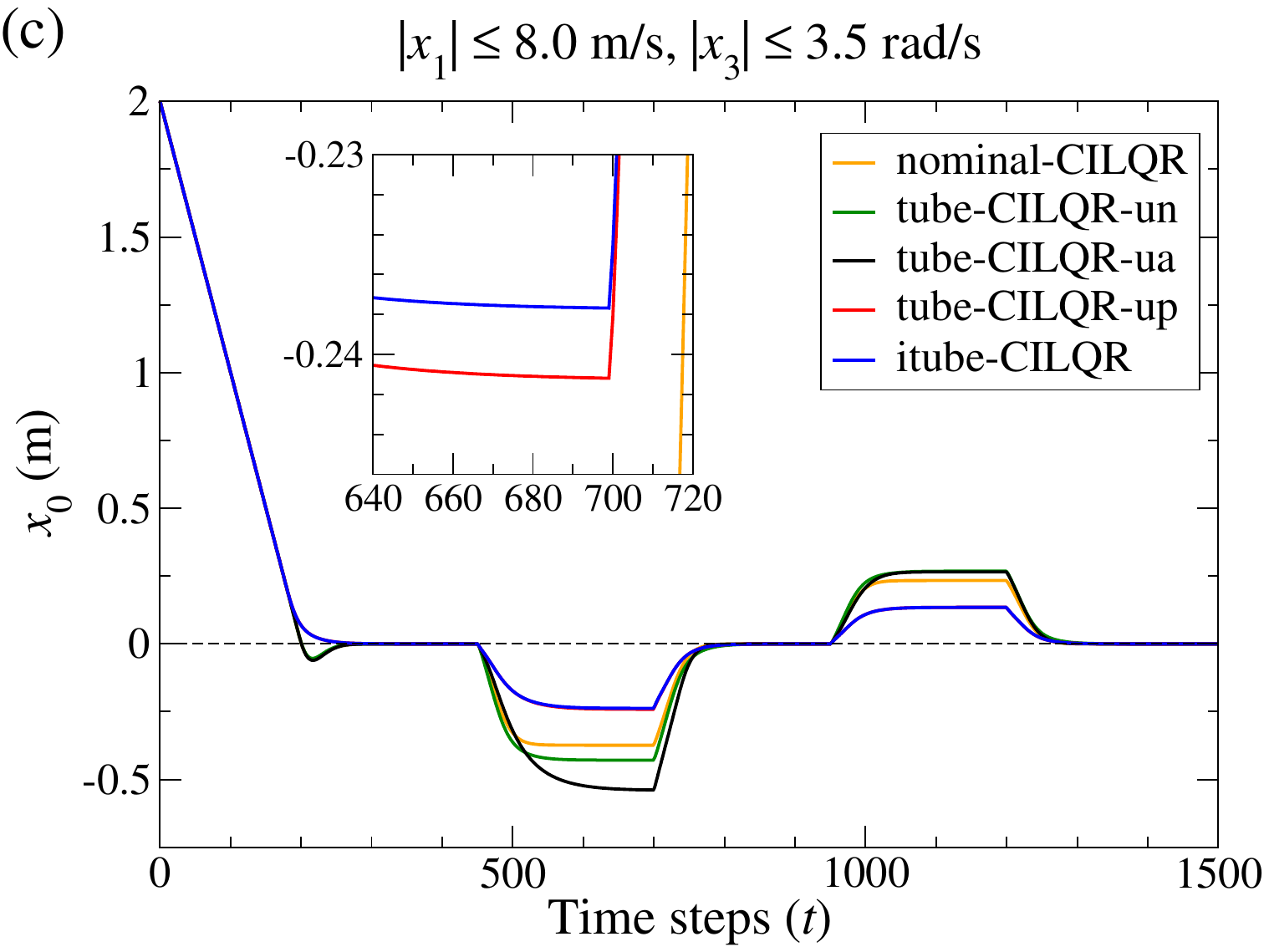}}%
{\includegraphics*[width=0.25\linewidth]{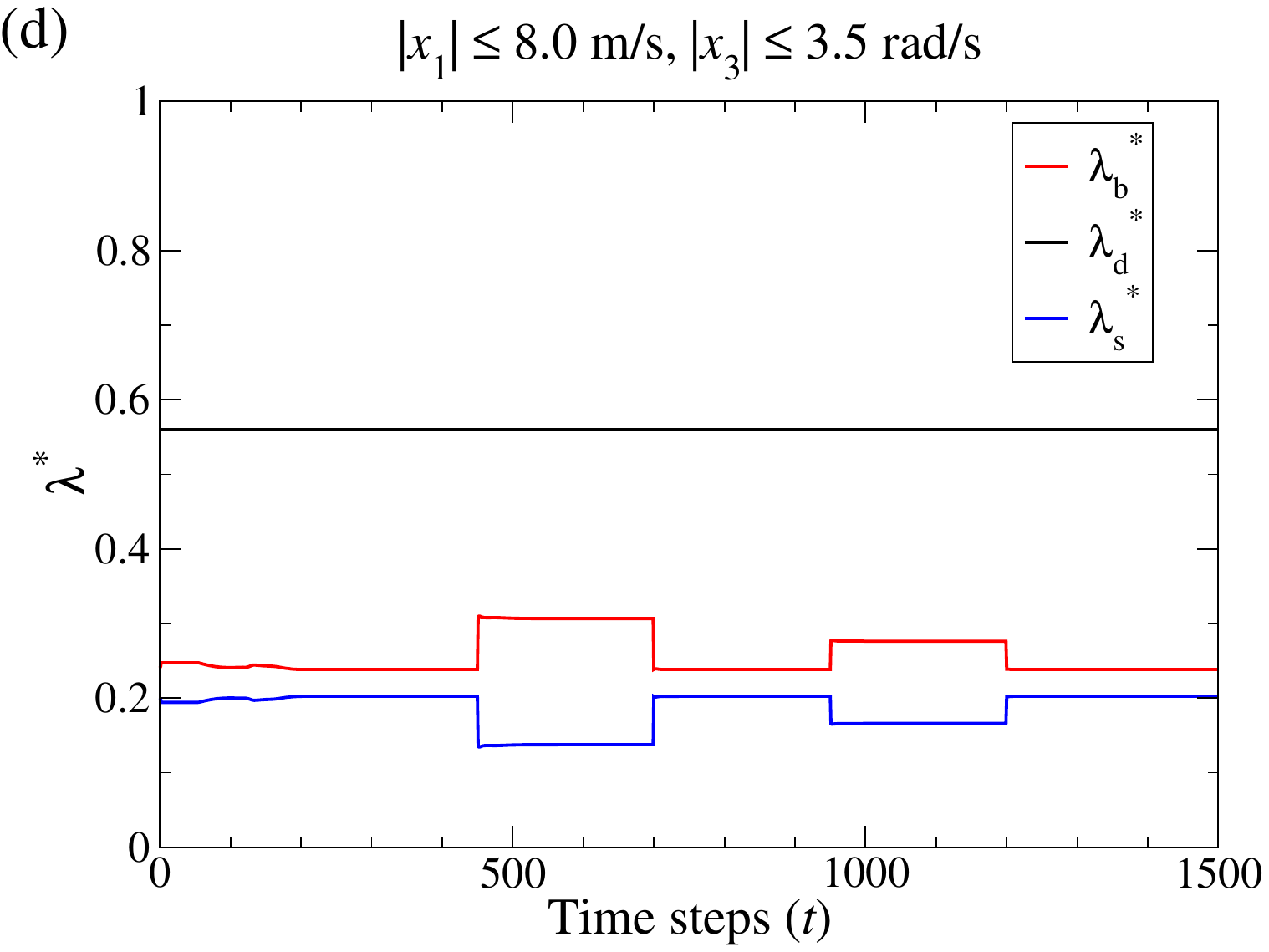}}%
\caption{Numerical simulation results obtained with different CILQR algorithms when $v_x$ = 20.0 m/s and $D$ = 0.22. Parts (a) and (b) were obtained when $\left| {x_1 } \right| \le$ 10.0 m/s and $\left| {x_3 } \right| \le$ 4.5 rad/s. Parts (c) and (d) were acquired when $\left| {x_1 } \right| \le$ 8.0 m/s and $\left| {x_3 } \right| \le$ 3.5 rad/s. In (a), the $x_0$ values obtained with the tube-CILQR-up and itube-CILQR algorithms at $t$ = 700 are -0.2078 and -0.2069 m, respectively. In (b), $\Delta \lambda  $ = 0.0136, 0.0821, and 0.0493 at $t$ = 0, 600, and 1100, respectively. In (c), the $x_0$ values obtained with the tube-CILQR-up and itube-CILQR algorithms at $t$ = 700 are -0.2411 and -0.2376 m, respectively. In (d), $\Delta \lambda  $ = 0.0402, 0.1691, and 0.1103 at $t$ = 0, 600, and 1100, respectively.} 
\end{figure*}

\section{Experimental Results and Discussion}
The proposed algorithm was evaluated against several other methods through numerical simulations and vision-based experiments in the TORCS environment. All simulations and experiments were conducted on a personal computer equipped with an Intel i9-9900K CPU, 64 GB of RAM, and an Nvidia RTX 2080 Ti graphics processing unit with  11 GB of VRAM. TORCS provides sophisticated visualization and physics engines and accurately simulates visual effects for a self-driving vehicle and its surrounding environment as well as vehicle dynamics \cite{Che15, Car15, Li-19, Lee21, Wu23}. Due to lag in the TORCS simulator, the actuation delay was approximately 6.66 ms on our computer for vision-based experiments; this lag was zero in numerical simulations. In addition, varying road adhesion levels were considered in the vision-based experiments, acting as external disturbances similar to sensor noise. Several remarks about the proposed vision-based autonomous driving scheme are provided in Sec. V-C.

\subsection{Numerical Simulations}
This section presents the numerical simulation results obtained with different CILQR- and MPC-based controllers. The test scenario involved an ego vehicle moving on a road with a cruising  speed of 20.0 m/s. The road had several straight sections and two turns with constant curvatures of $\kappa $ = 0.08 1/m (denoted as turn-1) and $\kappa $ = -0.05 1/m (denoted as turn-2), which were encountered at time steps 450 $\le$ $t$ $\le$ 700 and 950 $\le$ $t$ $\le$ 1200, respectively. The initial state of the ego vehicle was assumed to be ${\bf x}_0  = \left[ {\begin{array}{*{20}c}
   2 & 0 & 0 & 0  \\
\end{array}} \right]^T$. The following fixed settings were used in all simulations: $x_0  \equiv  \Delta $, $x_1  \equiv  {\dot \Delta } $, $x_2  \equiv  { \theta } $, $x_3 \equiv {\dot \theta }$, and  $u \equiv \delta$.

Fig. 4(a) displays the lateral offset ($ \Delta$) of the trajectories generated by the CILQR-based algorithms. The lateral offset represented the lateral position error of the ego vehicle with respect to the lane centerline at each time step, serving as a key performance indicator in the lane-keeping task. These trajectories exhibited plateau curves at turn-1 and turn-2, indicating an output offset attributable to the constant nonzero road curvature. Similar behavior was observed in the trajectory-following results for a  prototype autonomous four-wheel, independent-drive electric vehicle guided by an adaptive hierarchical control framework in a previous study (Fig. 19 in \cite{Guo18}). As displayed in Fig. 4(a), different CILQR-based algorithms exhibited a maximum lateral error at turn-1 near $t$ = 700. At this time, the tube-CILQR-un and tube-CILQR-ua algorithms exhibited larger errors than the other CILQR-based algorithms did. Moreover, these two algorithms exhibited lower performance than did the nominal CILQR algorithm because of the tightened constraints used in them. These constraints resulted in more conservative solution trajectories that deteriorated lateral control performance \cite{Wan22}. The tube-CILQR-up algorithm and the proposed itube-CILQR algorithm also outperformed the tube-CILQR-un and tube-CILQR-ua algorithms in terms of lateral control performance.

\begin{figure*}[!t]
\centering
{\includegraphics*[width=0.33\linewidth]{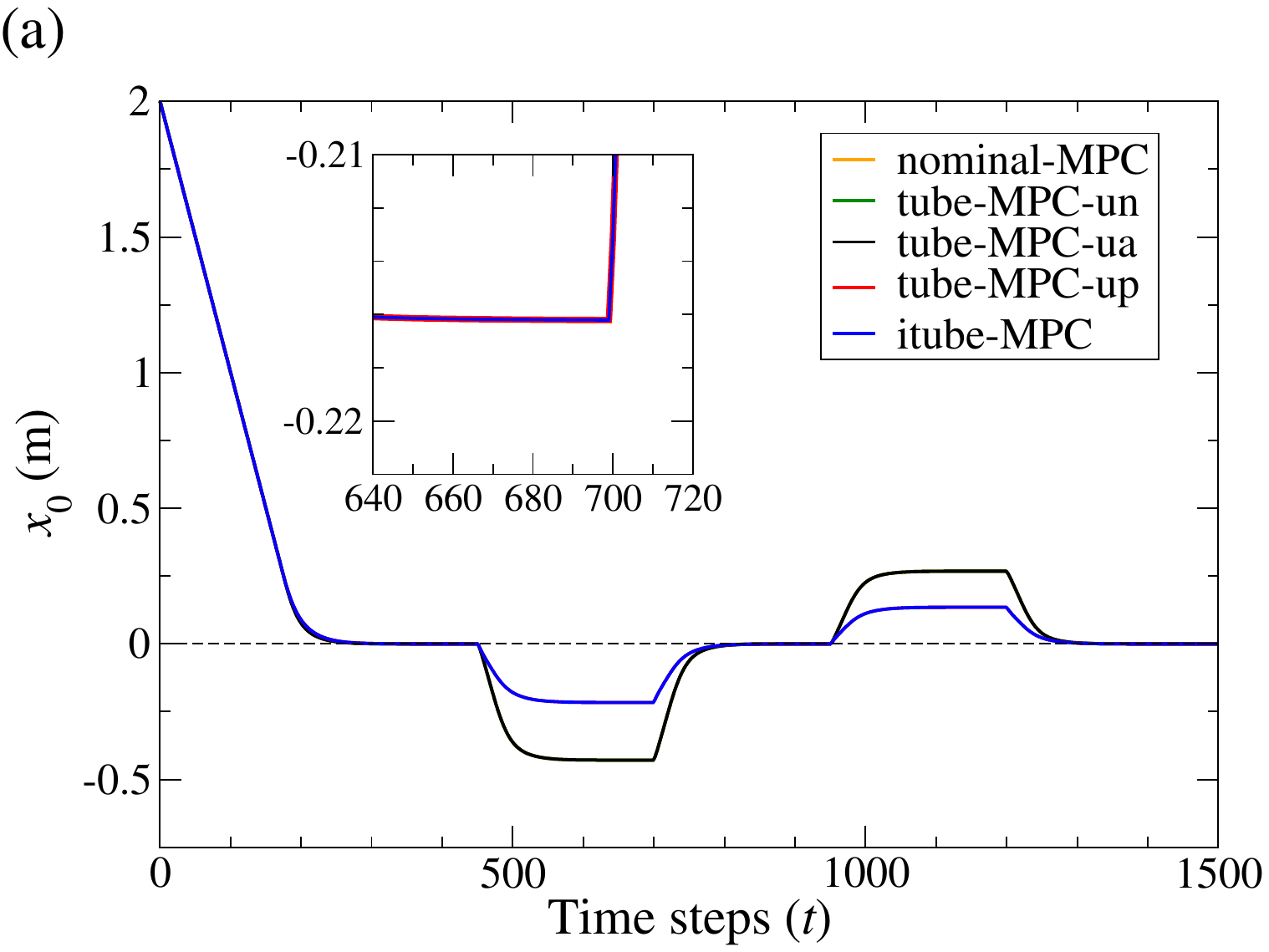}}%
{\includegraphics*[width=0.33\linewidth]{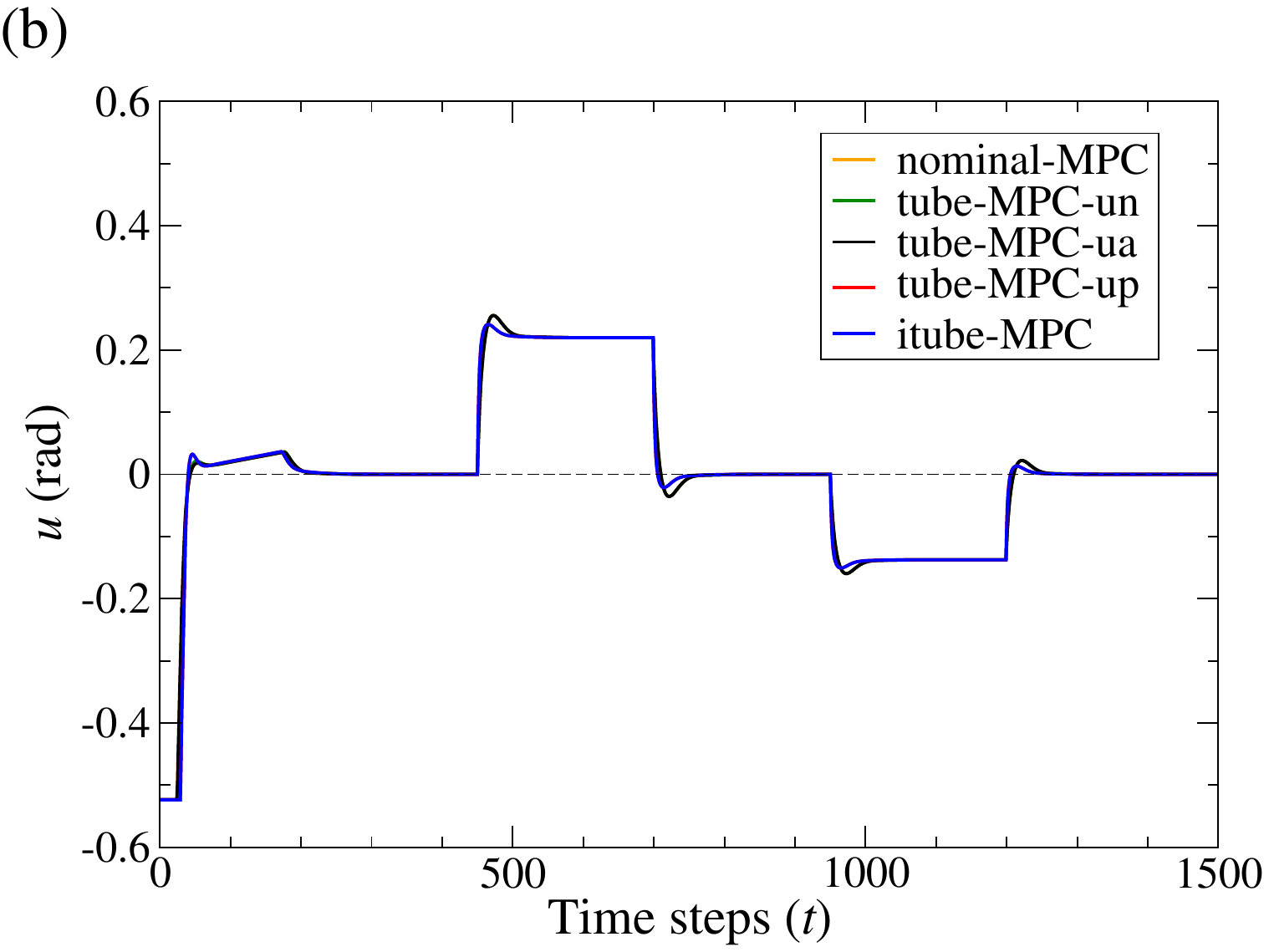}}%
{\includegraphics*[width=0.33\linewidth]{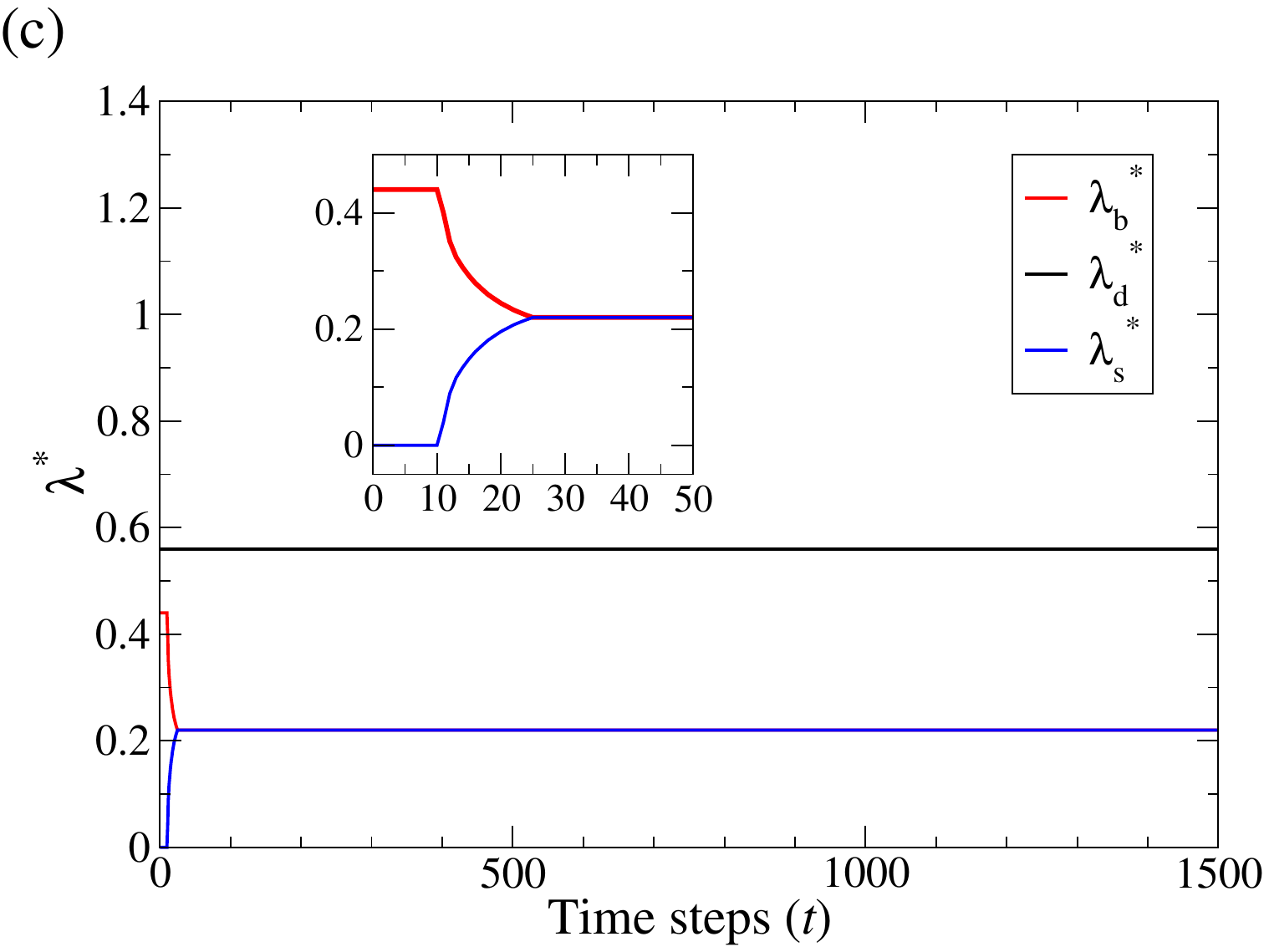}}%
\caption{Numerical simulation results obtained with different MPC algorithms when $v_x$ = 20.0 m/s, $D$ = 0.22, $\left| {x_1 } \right| \le$ 9.0 m/s, and $\left| {x_3 } \right| \le$ 4.0 rad/s. (a) Trajectory of the actual system state $x_0$. (b) Trajectory of the actual system control input. (c) Trajectories of the first components of the optimal interpolation variable sequences. In (a), the $x_0$ value obtained at $t$ = 700 with both the tube-MPC-up and for itube-MPC algorithms is -0.2162 m.} 
\end{figure*}

\begin{figure*}[!t]
{\includegraphics*[width=0.25\linewidth]{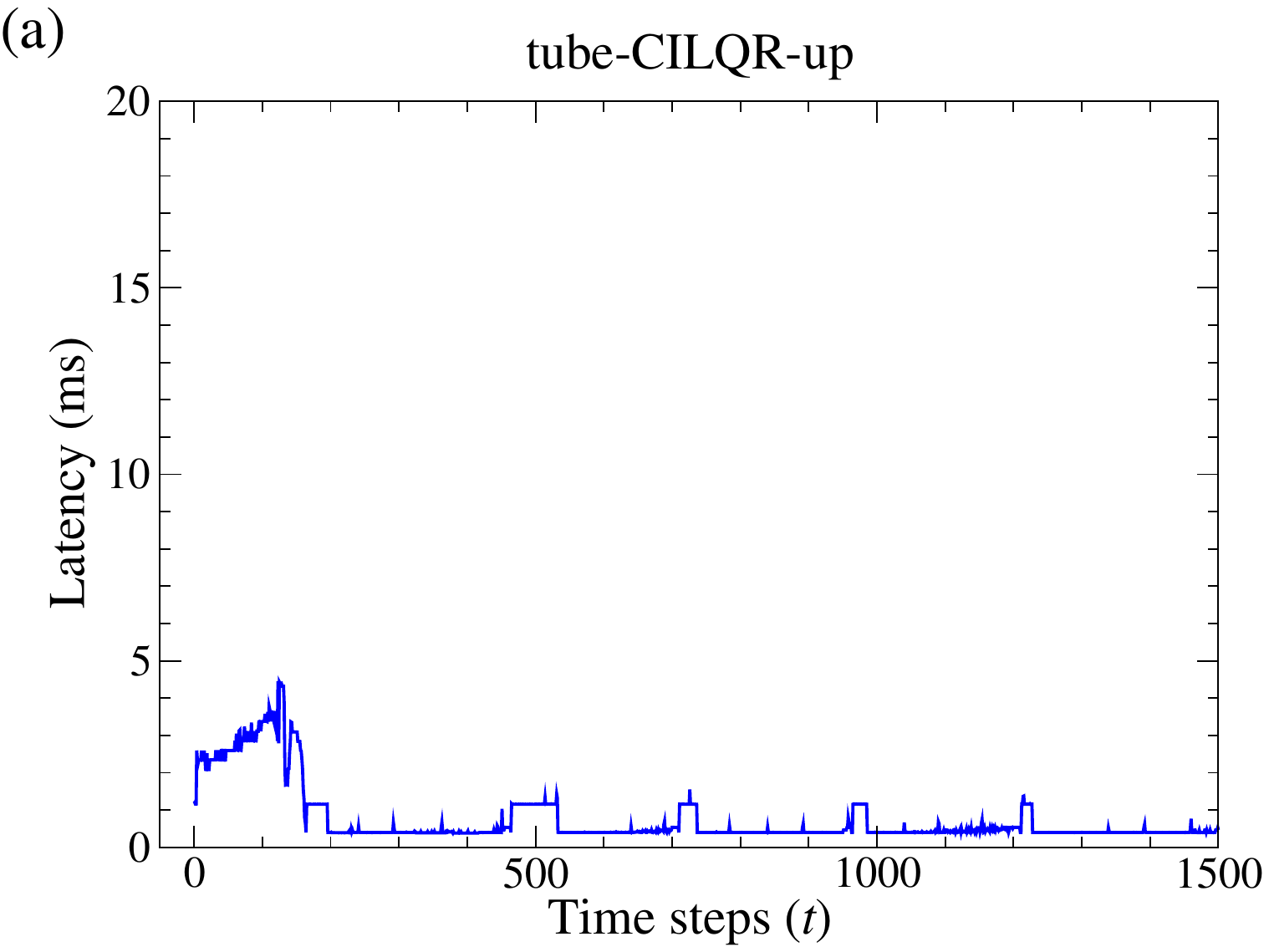}}%
{\includegraphics*[width=0.25\linewidth]{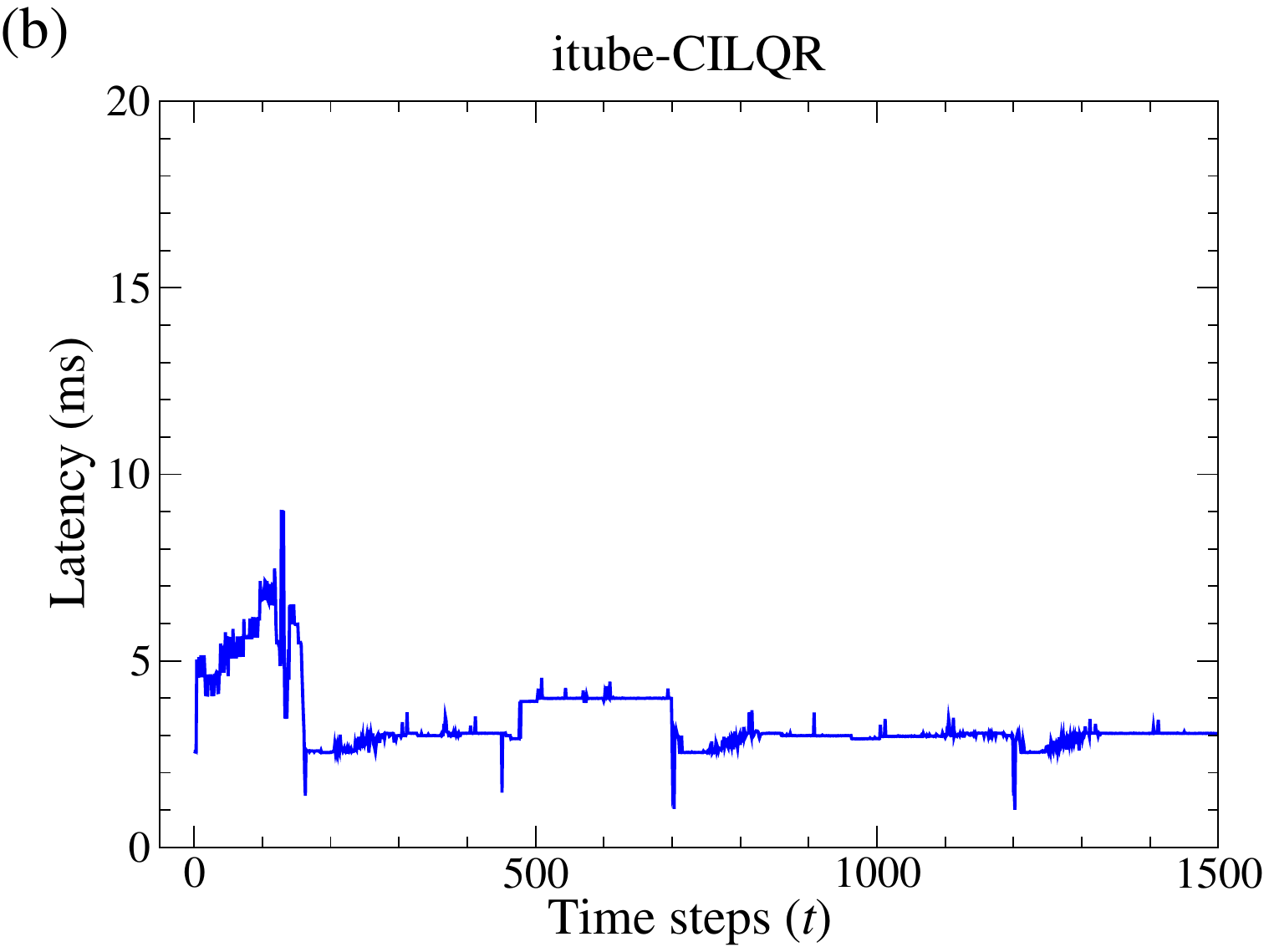}}%
{\includegraphics*[width=0.25\linewidth]{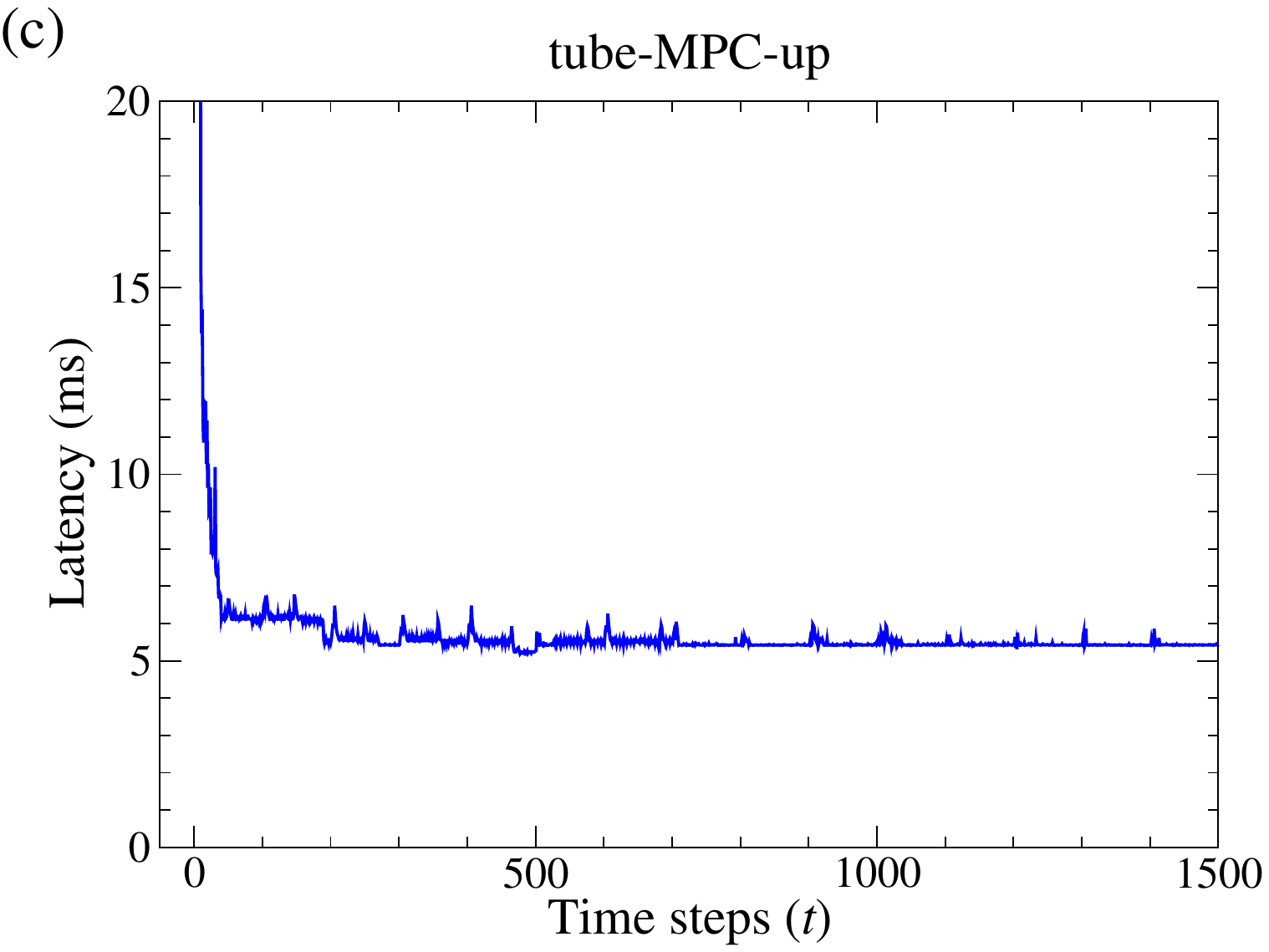}}%
{\includegraphics*[width=0.25\linewidth]{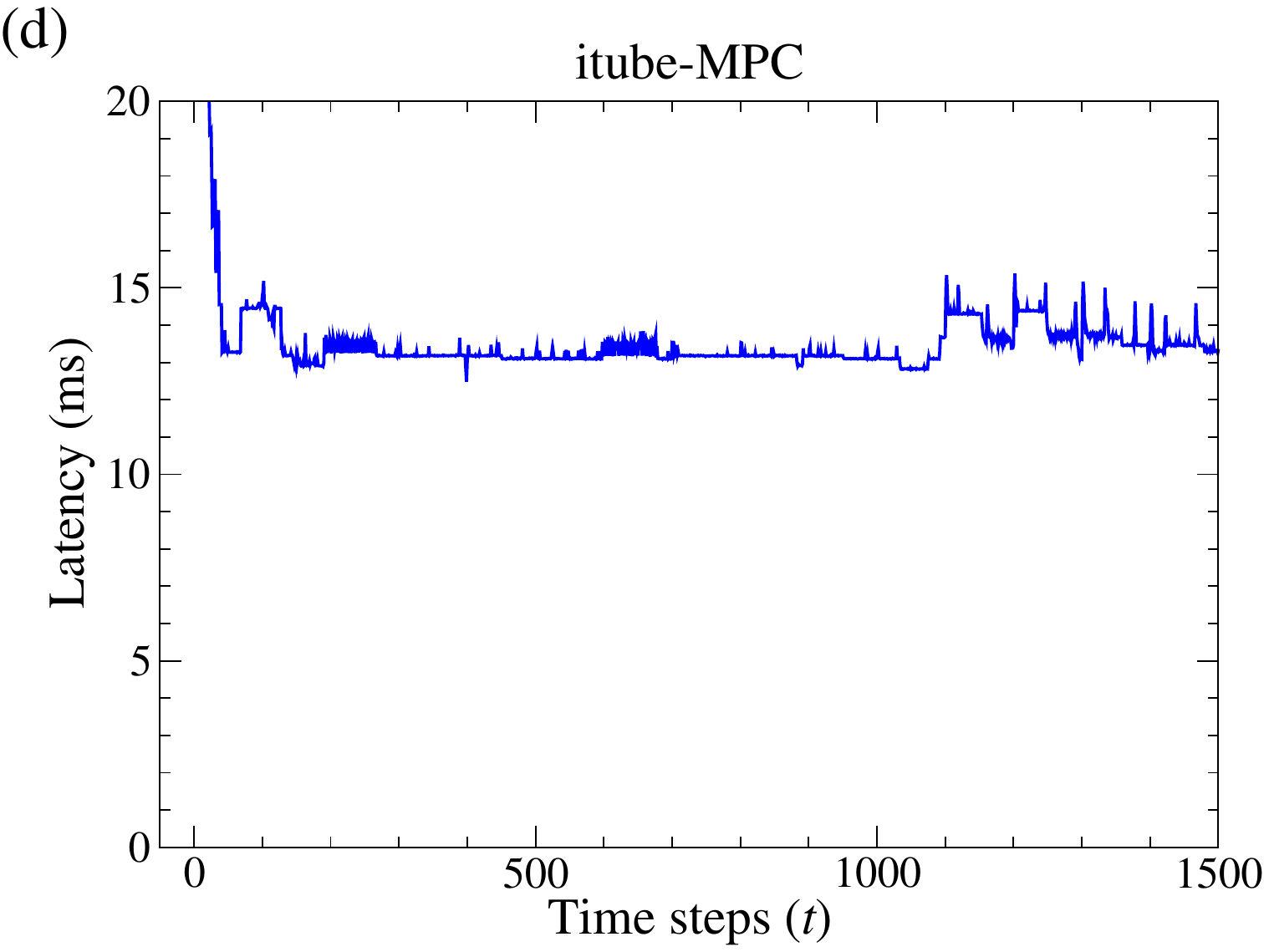}}%
\caption{Computation times of the (a) tube-CILQR-up, (b) itube-CILQR, (c) tube-MPC-up, and (d) itube-MPC algorithms at each time step.}
\end{figure*}

The tube-CILQR-up algorithm uses the control law  $u_p =  u_n + u_a$, enabling it to outperform the tube-CILQR-un and tube-CILQR-ua algorithms, which use the control law $u_n$ or $u_a$ alone, respectively. The advantage of the tube-CILQR-up algorithm is that the control laws $u_n$ and $u_a$ can collaboratively steer the $\Delta$-trajectory toward the target zero points. However, this control strategy may generate  nonsmooth control action trajectories when the control effort is high. As illustrated in Fig. 4(b), the  steering angle ($ \delta $) curve obtained with tube-CILQR-up was not smooth  at early time steps ($t$ $<$ 200). By contrast, the corresponding $\Delta$-trajectory displayed in Fig. 4(a) was stable at these time steps. The proposed itube-CILQR algorithm exhibited superior tracking performance when compared with the tube-CILQR-up algorithm. The itube-CILQR algorithm uses the integrated control law $u_p$ and applies interpolation to fuse multiple constraints. This method can reduce system conservatism and enhance controller performance. As illustrated in the subplot of Fig. 4(a), among all adopted CILQR-based methods, the itube-CILQR algorithm showed the lowest lateral error at turn-1 near $t$ = 700.

Fig. 4(c) displays the trajectories of the first element of the optimal interpolation variable sequences, namely ${\bf \Lambda }^*  = \left[ {\begin{array}{*{20}c}
   {\lambda _b^* } & {\lambda _d^* } & {\lambda _s^* }  \\
\end{array}} \right] ^T$. As described in Sec. IV, the parameter $\lambda _d^*$ was set to a constant value to reduce the computational effort. A standard method for comparing tube-MPC  methods in terms of conservatism involves measuring the size of the feasible domain  for  the considered optimization problems \cite{Kog20}. Instead of using this multidimensional approach, a one-dimensional quantitative analysis technique was employed in this study, with the conservatism of the system measured as the difference between $\lambda _b^*$ and $\lambda _s^*$ as follows:
\begin{equation}
\Delta \lambda  \equiv \lambda _b^*  - \lambda _s^* .
\end{equation}
Overall, the condition $\Delta \lambda  >$ 0  held during the entire maneuver, indicating that the less conservative system had a lower cost. Conservatism could be further reduced at the curves, where $\Delta \lambda$ increased with the road curvature. Moreover, the difference was larger at the turn with a larger curvature (turn-1), signifying that the trajectories of the interpolation variables could adaptively respond to the disturbance. A similar adaptive response behavior was observed in our previous work \cite{Lee23} for slack variables, which were incorporated into a CILQR algorithm to serve a similar role to that played by the interpolation variables employed in this study. The aforementioned adaptive response may be responsible for the itube-CILQR algorithm outperforming the tube-CILQR-up algorithm on the $\Delta $-trajectories [Fig. 4(a)]. Table III presents the experimental results obtained with the itube-CILQR algorithm at different $D$ values (13). Increasing $D$ (or decreasing $\lambda _d^*$) led to an increase in $\Delta \lambda $; however, itube-CILQR exhibited its lowest  lateral error near $D$ = 0.25. This sensitivity analysis enabled the identification of the working range of $D$.

\begin{table*}[!t]
\caption{Experimental results obtained with the itube-CILQR algorithm$^{a}$ under different $D$ values } 
\begin{center}
\begin{tabular}{c|c|c|c|c|c|c|c|c|c|c}\hline
$D$ &0.160 & 0.190  & 0.220 & 0.230 & 0.240& 0.250& 0.255& 0.258& 0.260& 0.270 \\ \hline
$\lambda _d^* $ & 0.680& 0.620& 0.560 &0.540&0.520 &0.500 &0.490 &0.484   &0.480   &0.460\\ \hline
$x_0$ $(t=700)$ &-0.2209 &  -0.2206 & -0.2201  & -0.2200 & -0.2198& -0.2197& -0.2200& -0.2207  & -0.2217  &-0.2220\\ \hline
$\Delta \lambda $ $(t=600)$&0.0926 & 0.1105  &  0.1283 &0.1341  &0.1399 &  0.1453 & 0.1480 &0.1496 &0.1507  &0.1561 \\  \hline
\multicolumn{11}{l}{$^{a}$\scriptsize{Parameters $v_x$ = 20.0 m/s, $\left| {x_1 } \right| \le$ 9.0 m/s, and $\left| {x_3 } \right| \le$ 4.0 rad/s were used.}} \\
\end{tabular}
\end{center}
\end{table*}

\begin{table*}[!t]
\caption{Average computation times for the tube-CILQR-up, itube-CILQR, tube-MPC-up, and itube-MPC algorithms$^{a}$}
\begin{center}
\begin{tabular}{c|c|c|c|c}\hline
Method &tube-CILQR-up &itube-CILQR & tube-MPC-up &  itube-MPC\\ \hline
Latency (ms)&0.76& 3.45 & 5.89 & 14.92 \\  \hline
\multicolumn{5}{l}{$^{a}$\scriptsize{All algorithms were implemented in C++.}} \\
\end{tabular}
\end{center}
\end{table*}

\begin{figure}[!t]
{\includegraphics*[width=0.5\linewidth]{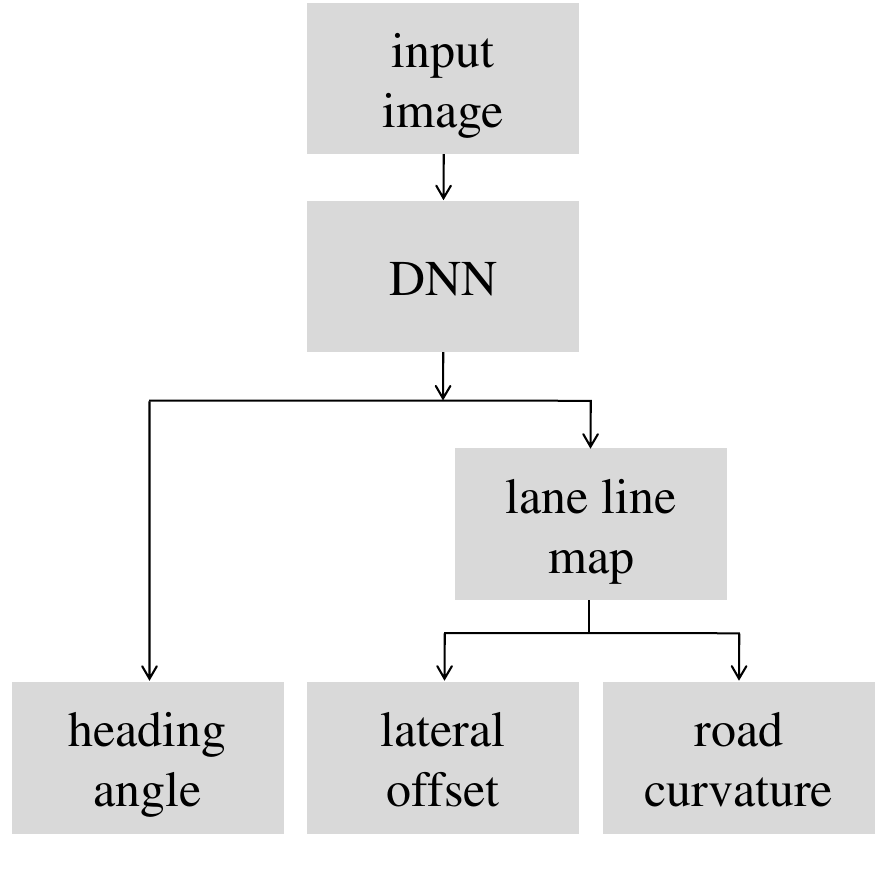}}%
{\includegraphics*[width=0.5\linewidth]{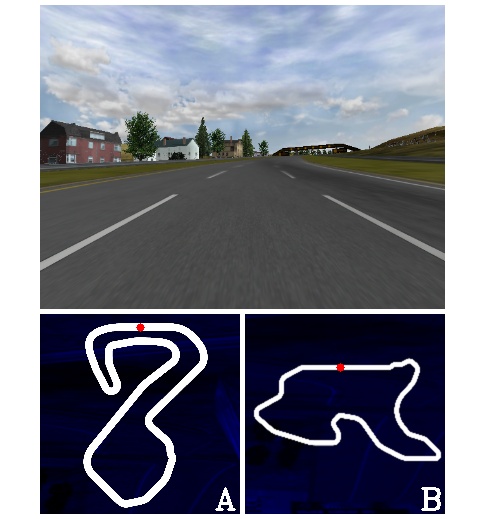}}%
\caption{Adopted vision perception system (left) \cite{Lee24}, and an example traffic situation and maps of track-A and track-B in TORCS for autonomous driving (right). The input of the adopted deep neural network model was an RGB image with a resolution of 228 $\times$ 228 pixels. The ego vehicle was designed to begin maneuvering at the locations marked by red symbols on the track maps, and the driving direction was counterclockwise. Both tracks contained several turns [Fig. 10(a) and Fig. 11(a)] and had different road friction values, which increased the difficulty of high-speed cornering. }
\end{figure}

Fig. 5 displays the lane-keeping results obtained for the CILQR algorithms when other constraint regions were used. The $\Delta$-trajectories in Fig. 4(a) and Fig. 5(a) and 5(c) reveal that the tracking performance improved (i.e., the lateral error decreased) when the system conservatism was decreased by increasing  the size of the constraint region. Consequently, at turn-1, itube-CILQR demonstrated the smallest lateral error when $\left| {x_1 } \right| \le$ 10.0 m/s and $\left| {x_3 } \right| \le$ 4.5 rad/s [Fig. 5(a)]. It exhibited larger lateral errors when $\left| {x_1 } \right| \le$ 9.0 m/s and $\left| {x_3 } \right| \le$ 4.0 rad/s [Fig. 4(a)] and when $\left| {x_1 } \right| \le$ 8.0 m/s and $\left| {x_3 } \right| \le$ 3.5 rad/s [Fig. 5(c)]. The $\lambda ^* $-trajectories in Fig. 4(c) and Fig. 5(b) and 5(d) indicate that the gap $\Delta \lambda$ generally increased when the size of the constraint region decreased; this trend was also observed for the difference in the lateral errors generated by itube-CILQR  and tube-CILQR-up at turn-1. At approximately $t$ = 700, the  largest lateral error difference was observed when $\left| {x_1 } \right| \le$ 8.0 m/s and $\left| {x_3 } \right| \le$ 3.5 rad/s [Fig. 5(c)].

The trajectories generated by different MPC-based controllers are presented in Fig. 6 for comparison. In general, the $\Delta$-trajectories produced by the MPC-based algorithms exhibited smaller lateral errors than did those generated by the CILQR-based algorithms [Fig. 6(a)]. This finding can be attributed to the fact that the MPC-based algorithms use the primal-dual-based IPOPT, whereas the CILQR-based algorithms employ simply primal-based optimization methods. The $\Delta$-trajectories in Fig. 6(a) can be classified into two groups: those produced by the nominal MPC,  tube-MPC-un, and tube-MPC-ua algorithms and those generated by the tube-MPC-up and itube-MPC algorithms. The tube-MPC-up and itube-MPC algorithms demonstrated smaller lateral errors than the other MPC-based methods did. Nevertheless, the $\Delta$-trajectories produced by the tube-MPC-up and itube-MPC algorithms exhibited similar trends, as indicated by the subfigure in Fig. 6(a). These trajectories were not split, in contrast to those produced by the tube-CILQR-up algorithm [Fig. 4(a)], even for a small constraint region with $\left| {x_1 } \right| \le$ 5.6 m/s and $\left| {x_3 } \right| \le$ 2.5 rad/s (not shown in the figures). Fig. 6(c) shows the $\lambda ^*$-trajectories derived using the itube-MPC algorithm. Compared  with the itube-CILQR algorithm [Fig. 4(c)], the itube-MPC algorithm initially exhibited lower conservatism (at $t$ $<$ 25) when the gap $\Delta \lambda$ was greater than 0. Subsequently, this gap converged to 0, even at the curves. These results may explain why the itube-MPC and tube-MPC-up algorithms generated similar $\Delta $-trajectories at turn-1 and turn-2 [Fig. 6(a)]. The IPOPT likely produced near-optimal solutions for both of these algorithms, leading to them generating similar results.

\begin{figure*}[!t]
\centering
{\includegraphics*[width=0.33\linewidth]{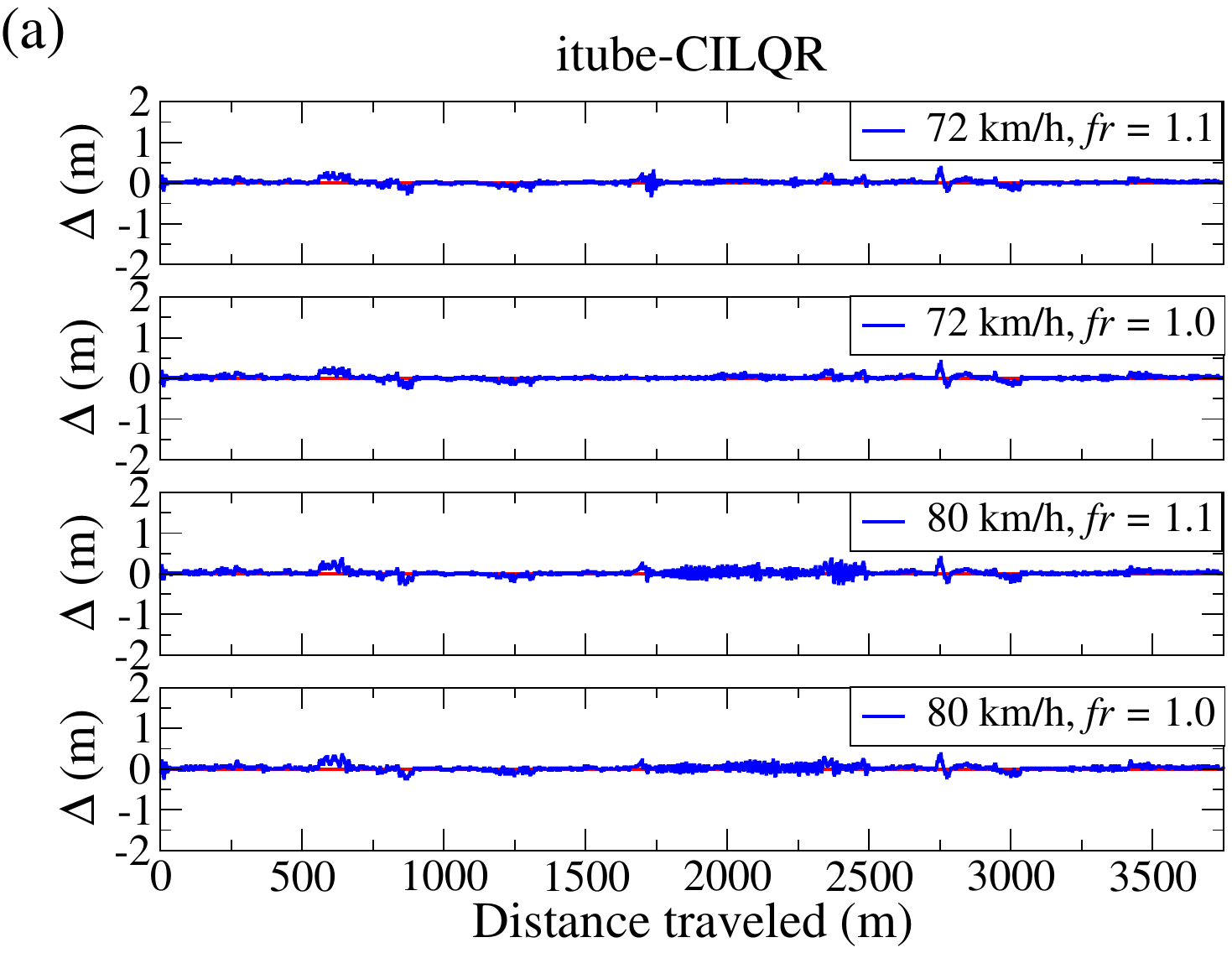}}%
{\includegraphics*[width=0.33\linewidth]{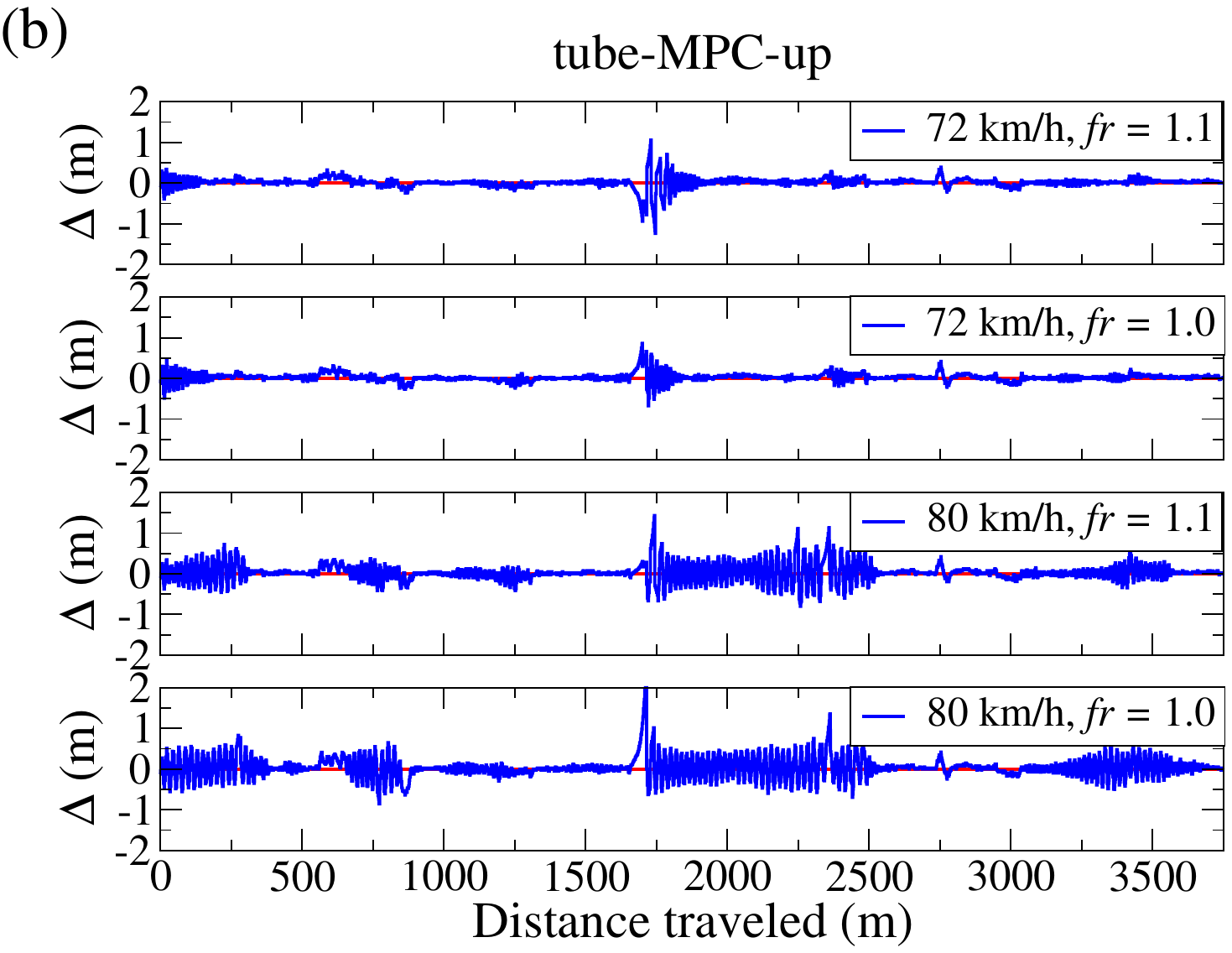}}%
{\includegraphics*[width=0.33\linewidth]{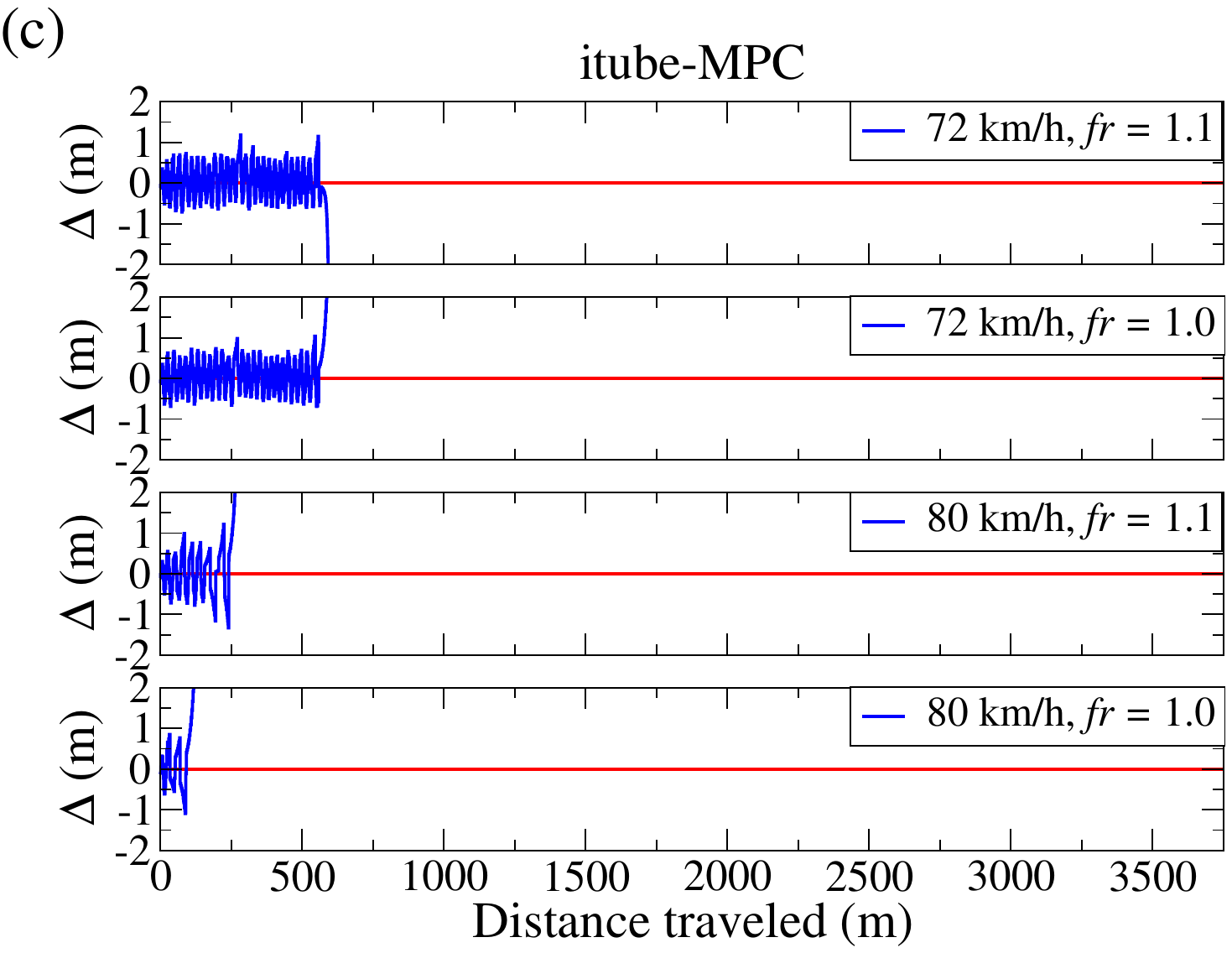}}%
\caption{Trajectories obtained for $\Delta$ by using the (a) itube-CILQR, (b) tube-MPC-up, and (c) itube-MPC algorithms when $fr$ = 1.1 or 1.0 for an ego vehicle traveling on track-A at 72 or 80 km/h.} 
\end{figure*}

The computational times required by the tube-CILQR-up, itube-CILQR, tube-MPC-up, and itube-MPC algorithms to generate a trajectory are displayed in Fig. 7(a), 7(b), 7(c), and 7(d), respectively. At early time steps ($t$ $<$ 200), the tube-CILQR-up and itube-CILQR algorithms required up to 5 and 10 ms, respectively, to generate a solution. Compared with these algorithms, the MPC-based algorithms required a longer time ($>$20 ms) to generate solutions. This result can be attributed to the difficulty exhibited by the IPOPT in finding appropriate starting points for optimization owing to its reliance on the interior-point method \cite{Wac06}. On average, compared with itube-CILQR, the tube-MPC-up and itube-MPC algorithms required approximately 1.71 and 4.32 times longer computational times, respectively, to generate control commands (Table IV). Therefore, a clear trade-off between computational burden and performance was observed for the itube-CILQR and itube-MPC algorithms. Compared with itube-MPC, itube-CILQR had a lower computational cost. Moreover, itube-CILQR maintained satisfactory performance, exhibiting a lateral error of less than 0.23 m at turn-1 [Fig. 4(a)]. In accordance with these findings, the proposed itube-CILQR algorithm with the parameter setting in Fig. 4 was compared with its counterparts, the tube-MPC-up and itube-MPC algorithms, for vision-based maneuvers. The results are in the following section.

\subsection{Vision-Guided Lane-Keeping Experiments}
Vehicle lane-keeping control is a key function of autonomous vehicles, and numerous researchers have employed computer-vision-based approaches to evaluate this function \cite{Xu20, Get24, Kum22, Chung24, Siz25}. A vision system developed in our previous work \cite{Lee24}, which uses a deep neural network (DNN) model to extract image features, was employed in the present study along with control algorithms to maneuver an ego vehicle in a simulated environment within TORCS. The DNN model used in the visual perception module could output data for autonomous driving at a speed of 40 frames per second (Fig. 8). It  simultaneously generated the ego vehicle's heading angle ($\theta$) and a binary lane map by using a multitask UNet architecture (MTUNet) \cite{Lee24, Ron15}. The lateral offset $\Delta$ and road curvature $\kappa$ were then derived using postprocessing methods \cite{Lee24}. The terms $\theta$, $\Delta $, and $\kappa$ were obtained with respect to the current position of the ego vehicle. Subsequently, the itube-CILQR, tube-MPC-up, and itube-MPC algorithms used these parameters and zero derivatives (${\dot \Delta }$ and ${\dot \theta }$) as inputs to guide the ego vehicle to track the lane centerline in TORCS \cite{Lee24, Lee23}. The tube-MPC-up algorithm operated under tightened constraints, which were looked up directly from the $\kappa$-table described in Sec. III-D. The itube-CILQR and itube-MPC algorithms used the tightened constraints from the $\kappa$-table to compute improved tightened constraints in accordance with the procedure described in Sec. III-C. The test environments, namely track-A and track-B, were three-lane highways with a lane width of 4 m (Fig. 8), and their total lengths were 3919 and 4441 m, respectively, and maximum curvatures were approximately 0.05 and 0.03 1/m, respectively. The road friction ($fr$) in the simulated environment was varied within the range of [0.8, 1.1] to test the robustness of the maneuvering performance \cite{Bon17}. The DNN model was trained on data for track-B but not on data for track-A to simulate unsupervised and supervised training scenarios. Vehicle maneuvering was successful in both cases \cite{Wu23}. In addition to itube-CILQR, tube-MPC-up, or itube-MPC (lateral control), a  proportional\textendash integral controller \cite{Lee24} was used to control the ego vehicle so that it maintained a constant longitudinal speed of 72 or 80 km/h (20.0 or 22.2 m/s, respectively) during lane-keeping maneuvers.

\begin{figure}[!t]
\centerline{\includegraphics[scale=0.33]{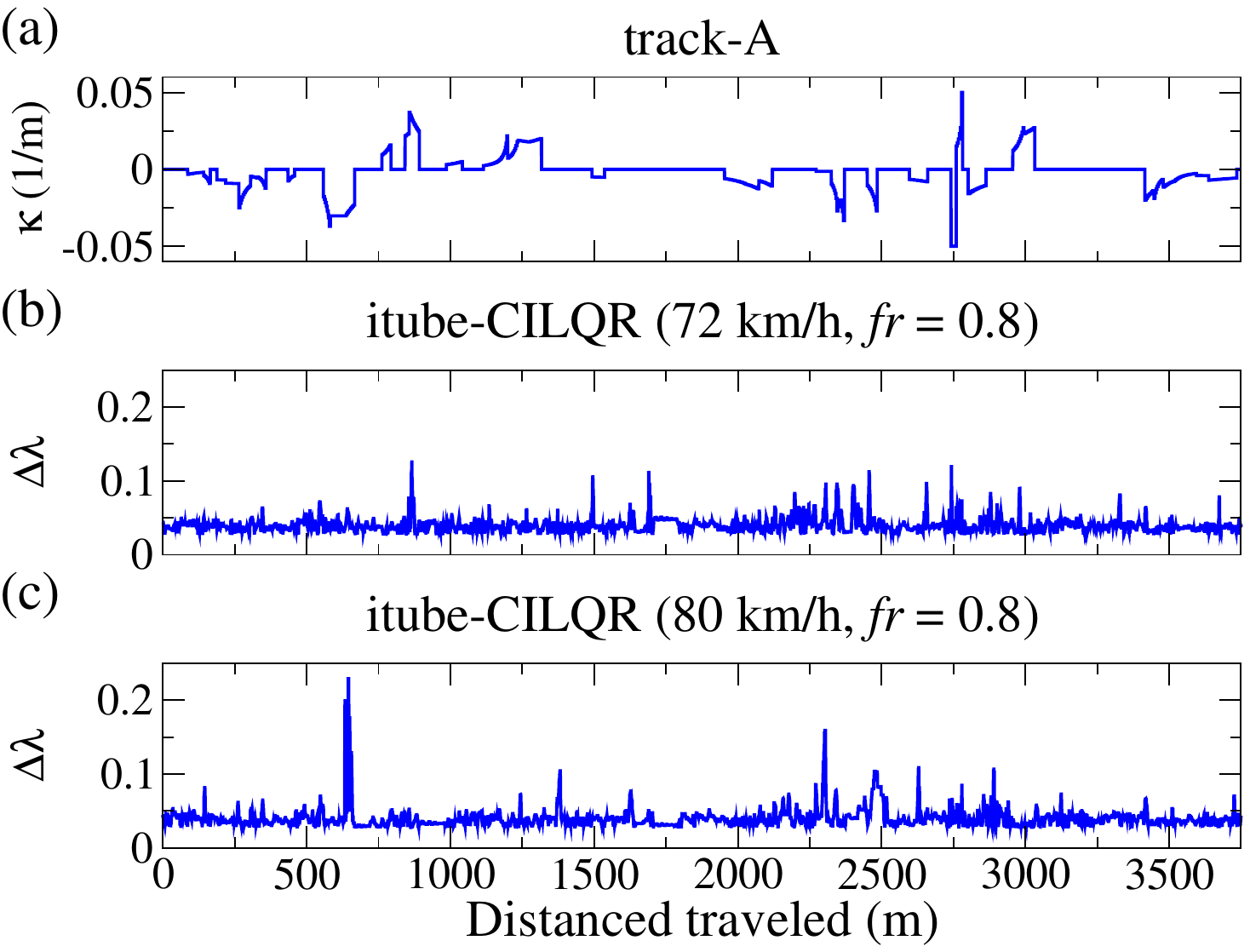}}
\caption{Lane curvature results ($\kappa$) and computed gaps ($\Delta \lambda$) for track-A. (a) Curvature profile of track-A. Trajectories obtained for $\Delta \lambda$ by using the itube-CILQR algorithm when $fr$ was 0.8 and the ego vehicle was traveling  at (b) 72 and (c) 80 km/h. The results of the associated quantitative analyses are presented in Table V.}
\end{figure}

\begin{figure}[!t]
\centerline{\includegraphics[scale=0.33]{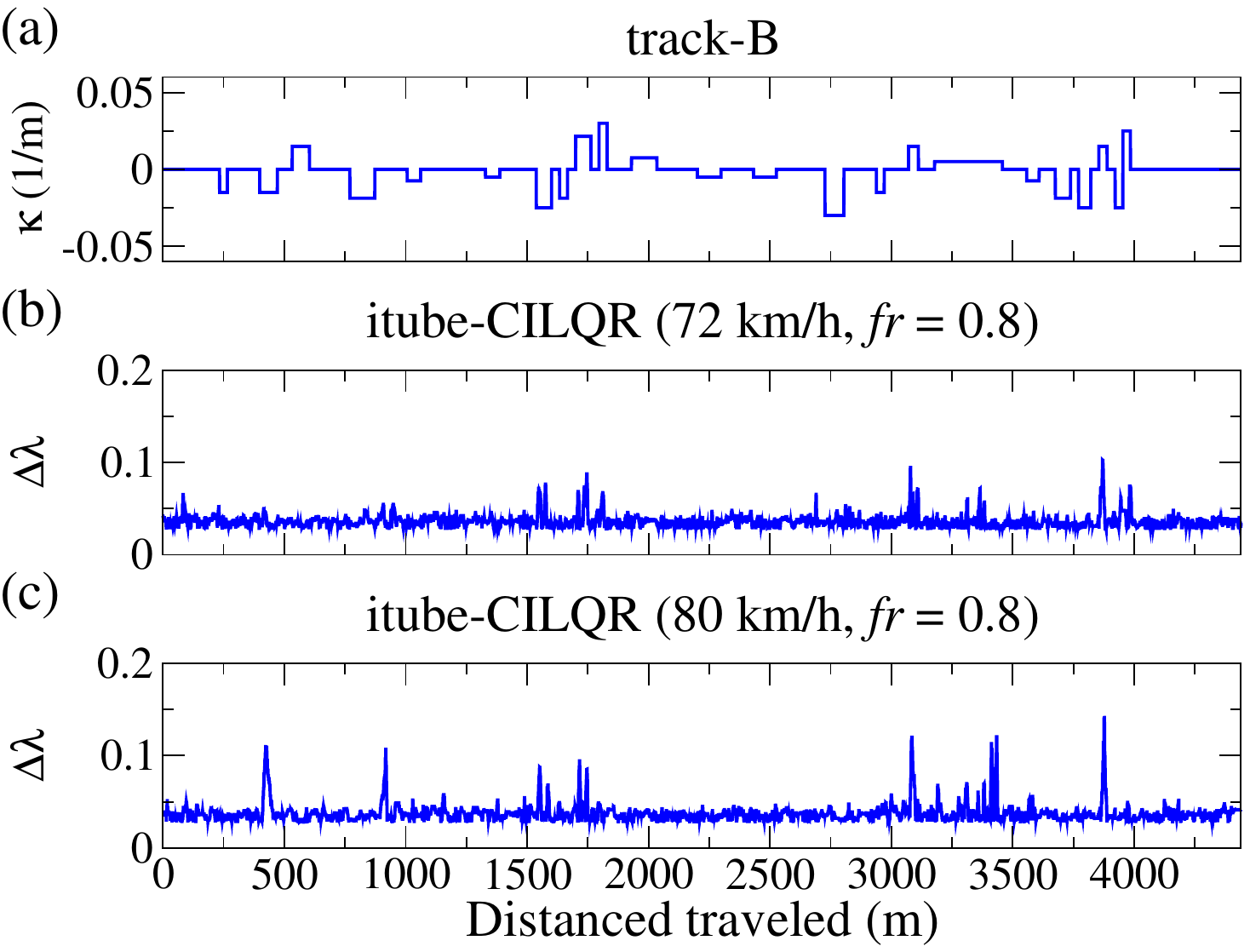}}
\caption{Lane curvature results ($\kappa$) and computed gaps ($\Delta \lambda$) for track-B. (a) Curvature profile of track-B. Trajectories obtained for $\Delta \lambda$ by using the itube-CILQR algorithm when $fr$ was 0.8 and the ego vehicle was traveling at (b) 72 and (c) 80 km/h. The results of the associated quantitative analyses are presented in Table V.}
\end{figure}

Fig. 9 depicts the $\Delta$-trajectories generated for the ego vehicle by the itube-CILQR, tube-MPC-up, and itube-MPC algorithms at a cruising speed of 72 or 80 km/h on track-A when $fr$ = 1.0 or 1.1. The itube-CILQR and tube-MPC-up algorithms could appropriately maneuver the ego vehicle, which accomplished the lane-keeping task over a complete lap on track-A. The ego vehicle remained closer to the lane center (i.e., it exhibited a lower lateral error) when it was controlled by the proposed itube-CILQR algorithm than when it was controlled by the tube-MPC-up algorithm. The  itube-CILQR algorithm showed a lateral error of less than 0.5 m in all cases. However, the $\Delta$-trajectories produced by the tube-MPC-up algorithm oscillated substantially, resulting in a larger average lateral error. This phenomenon resulted from steering lag, which was caused by finite controller latency \cite{Peng21}. The  latency of itube-CILQR was shorter than the actuation delay (6.66 ms) of the ego vehicle in TORCS, whereas the latency of tube-MPC-up generally equaled or exceeded this delay. Moreover, because of the long latency of the itube-MPC algorithm, the trajectories generated by it exhibited larger oscillations than did those produced by tube-MPC-up. Thus, the ego vehicle deviated substantially from the lane center when it was controlled by itube-MPC. In the worst case, itube-MPC failed to keep the ego vehicle within the lane at a distance of approximately 120 m at a cruising speed of 80 km/h and $fr$ = 1.0. In summary, the itube-CILQR algorithm outperformed the tube-MPC-up and itube-MPC algorithms in vision-guided lane-keeping maneuvers, which can primarily be attributed to its higher computation speed.

\begin{table*}[!t]
\caption{Average and standard deviation values of $\Delta \lambda$ during lane-keeping maneuvers performed using the itube-CILQR algorithm}
\begin{center}
\begin{tabular}{c|c|c|c|c|c|c|c|c}\hline
\multirow{3}{*}{$fr$}& \multicolumn{4}{c|}{track-A}   & \multicolumn{4}{c}{track-B}   \\ \cline{2-9}
 & \multicolumn{2}{c|}{80 km/h} &\multicolumn{2}{c|}{72 km/h} &\multicolumn{2}{c|}{80 km/h}& \multicolumn{2}{c}{72 km/h} \\ \cline{2-9}
  & Avg & SD & Avg & SD & Avg & SD & Avg & SD \\ \hline
0.80&0.040587&0.013882&0.040110&0.009718&0.037401&0.009966&0.036320&0.006903
\\ \hline

0.85&0.040086&0.009908&0.039750&0.009818&0.037047&0.008608&0.036401&0.007932
\\ \hline

0.90&0.039868&0.009830&0.039938&0.009995&0.036809&0.008079&0.036516&0.008138
\\ \hline

1.00&0.040076&0.010097&0.039445&0.008625&0.036859&0.009265&0.036303&0.007021
\\ \hline

1.10&0.040528&0.010374&0.039603&0.009133&0.036799&0.008443&0.036166&0.006367
\\ \hline

Avg&0.040229&0.010818&0.039769&0.009458&0.036983&0.008872&0.036341&0.007272
\\ \hline
\end{tabular}
\end{center}
\end{table*}

Figs. 10 and 11 display the curvature profiles and $\Delta \lambda $-trajectories obtained for track-A and track-B, respectively, under different conditions by using the itube-CILQR algorithm. In general, the $\Delta \lambda $ values were greater than 0 for both tracks, indicating that itube-CILQR was less conservative than standard tube-based approaches. The peaks in the $\Delta \lambda $-trajectories correspond to predicted cornering events involving large road curvature. This trend was generally similar to that observed in the numerical simulations [Fig. 4(c) and Fig. 5(b) and 5(d)]. The results derived for track-A (unsupervised learning) were noisy [Fig. 10(b) and 10(c)], which can be attributed to several factors, such as the irregular curvature profile, algorithm latency, the insufficient detection range of the lane detector \cite{Xu20}, the unstable and inconsistent lane inference results \cite{Get24}, and poor lane-marker quality \cite{Chung24}. These general problems can reduce safety for autonomous vehicles that use lane detection results. The $\Delta \lambda$-trajectories for maneuvers performed on track-B (supervised learning) showed fewer distinct peaks than did those for maneuvers performed on track-A [Fig. 11(b) and 11(c)]. The corresponding quantitative analyses are presented in Table V. The average and standard deviation values of $\Delta \lambda$ were lower for track-B than for track-A, signifying that superior vehicle maneuvering results are achieved for roads included in the training process than for unseen roads. Thus, training the adopted DNN model on the target road is a straightforward approach for improving the lateral control performance of itube-CILQR, thereby enhancing the safety of autonomous driving.

\subsection{Remarks on the Proposed Vision-Based Autonomous Driving Scheme}
Some remarks related to the experiments described in Sec. V are provided as follows. First, a self-driving vehicle guided by the proposed itube-CILQR control algorithm exhibited excellent driving performance in high-speed cornering scenarios within a TORCS environment. In this environment, sensing was conducted using a front-facing camera, and MTUNet \cite{Lee24} was used to collect and analyze real-time data for vision-only lane-keeping maneuvers, with a focus on high-speed cornering; thus, the controller inputs (${\Delta  }$ and ${ \theta}$) were derived through visual sensing. This fast approach enables the control target to be optimized and maneuver-associated noise (arising from aspects such as rapid changes in the detected road curvature, the discretization of the $\kappa$-table, or the use of simplified linear dynamics at tight turns) to be suppressed through the optimization of auxiliary interpolation variables; thus, it can achieve a suitable balance between control accuracy and computational efficiency.

Second, this study used zero state derivatives (${\dot \Delta }$ and ${\dot \theta }$) as partial controller inputs, which may affect the applicability of the proposed algorithm. The parameter ${\dot \theta }$ can be obtained using an inertial measurement unit (IMU) sensor, and ${\dot \Delta }$ can then be estimated through the Kalman filter \cite{Lev22}. This approach is general and can effectively enhance the real-world performance of vision-only autonomous driving systems.

Third, real vehicle tests can be conducted to evaluate the proposed algorithms. However, vision-based lane-keeping experiments for real vehicles are often restricted to low speeds or small road curvatures because of constraints related to safety, testing duration, and cost. For example, a speed of 5\textendash15 mph (8.04\textendash24.14 km/h) was used in \cite{Get24}, and a curvature smaller than 0.001 1/m was used in \cite{Chung24}. Thus, the conditions considered in this study (Tables I and II) are difficult to replicate in real-world tests. Hardware-in-the-loop (HIL) tests provide an easier approach for validating system performance before actual automotive implementation \cite{hil04, hil20}. In these tests, a real vehicle is secured to a mechanical device and driven in a virtual environment, allowing researchers to safely assess complex advanced driver assistance systems under various challenging conditions at low cost. State-of-the-art machine learning techniques can improve system performance in HIL tests \cite{hil20}. In our previous study \cite{Lee24}, MTUNet was trained on a synthetic TORCS dataset in which image data were annotated with the ego vehicle’s heading angle, the road type, ego-lane lines, and traffic object bounding boxes. However, to the best of our knowledge, no single real-world dataset contains all of these labels, representing a major barrier to automotive advancements.

\section{Conclusion}
To improve vehicle control and safety in autonomous driving, this study developed the itube-CILQR algorithm, which integrates the CILQR optimization algorithm \cite{Che19, Liu24} with an interpolation-tube-based MPC approach \cite{Kog20}. The CILQR algorithm provides high computation speed, and the interpolation-tube-based method enables the computation of improved tightened constraints through varying disturbance bounds. Furthermore, the proposed itube-CILQR algorithm uses an integrated control law \cite{Dav11} to achieve robust control of the actual system. The  disturbance considered in this study resulted from road curvature, which could be detected online along with the ego vehicle's heading angle and the lateral offset during a lane-keeping task by using a visual perception system \cite{Lee24}. The effectiveness of the proposed algorithm was validated through numerical simulations and vision-based experiments conducted in TORCS for large-curvature roads with different friction levels. The itube-CILQR algorithm exhibited satisfactory performance in the numerical simulations and outperformed the itube-MPC algorithm, which uses the IPOPT, in real-time vision-based maneuvers. The itube-CILQR algorithm had a short average computation time of approximately 3.45 ms in the vision-based experiments, whereas the itube-MPC algorithm had a relatively long average computation time of 14.92 ms. Moreover, compared with standard tube-based approaches, such as tube-CILQR-up, itube-CILQR reduced conservatism, thereby enhancing system control performance.

Overall, this work represents a step toward robust vision-guided automotive control systems. Future studies can focus on two directions. First, they can investigate whether the findings of the present study hold true when additional factors related to automated driving are considered, such as the interactions of the ego vehicle with surrounding vehicles or emergency obstacle avoidance. Second, they can examine the effectiveness of the proposed algorithm through physical experiments on HIL or actual vehicle platforms.

%authors’ photography is unnecessary, and authors’ biography is provided below.

%received his Ph.D. in Physics from the Department of Electrophysics, %National Yang Ming Chiao Tung University, Taiwan, in 2018. His research %interests include autonomous driving, computer vision, and machine %intelligence.
%\end{IEEEbiographynophoto}

\end{document}